\documentclass[journal]{IEEEtran}
\IEEEoverridecommandlockouts

\usepackage{array}
\usepackage{bbm}
\usepackage[labelsep=period,font=small]{caption}
\usepackage[subrefformat=parens,farskip=0pt,justification=centering]{subfig}
\usepackage{balance}
\usepackage{graphicx}
\usepackage{authblk}
\usepackage{booktabs}
\usepackage{textcomp}   
\usepackage{stfloats}
\usepackage{url}
\usepackage{verbatim}
\usepackage{graphicx}
\usepackage{cite}
\usepackage{makecell}
\usepackage{multirow}
\usepackage{multicol}
\usepackage{comment}
\usepackage{textcomp}
\usepackage{enumitem}
\usepackage{threeparttable}
\usepackage{physunits}
\usepackage[round-pad=false, round-mode=places,round-precision=2,group-separator={,},output-decimal-marker={.}]{siunitx}
\usepackage{pgfplots}
\pgfplotsset{compat=newest}
\usepgfplotslibrary{groupplots}

\usepackage{etoolbox}
\makeatletter
\patchcmd{\@algocf@start}{-1.5em}{0pt}{}{} 
\newif\if@restonecol
\makeatother

\usepackage[noend,linesnumbered,ruled,vlined]{algorithm2e}
\usepackage{algpseudocode}
\usepackage{amsmath,amssymb,amsfonts}
\SetKwRepeat{Do}{do}{while}
\SetAlCapHSkip{0pt} 
\allowdisplaybreaks

\usepackage[bookmarks=false]{hyperref}
\hypersetup{
	colorlinks=true,
	linkcolor=blue,
	filecolor=black,      
	urlcolor=teal,
	citecolor=blue,
}

\usepackage{pifont}
\usepackage{colortbl}
\usepackage{listings}

\newcommand{\cmark}{\ding{51}}%

\begin{document}

\twocolumn

\title{Towards Secure and Efficient DNN Accelerators via Hardware-Software Co-Design}

\author{\IEEEauthorblockN{Wei Xuan, Zihao Xuan, Rongliang Fu, Ning Lin, Kwunhang Wong, Zikang Yuan, Lang Feng, Zhongrui Wang, Tsung-Yi Ho~\IEEEmembership{Fellow,~IEEE}, Yuzhong Jiao, and Luhong Liang}
\thanks{The research was partially supported by the National Key R\&D Program of China (Grant No. 2022YFB3608300). This research was partially conducted by ACCESS – AI Chip Center for Emerging Smart Systems, supported by the InnoHK initiative of the Innovation and Technology Commission of the Hong Kong Special Administrative Region Government. It was also supported in part by Shenzhen Science and Technology Innovation Commission (Grant No. SGDX20220530111405040), Beijing Natural Science Foundation (Grant No. Z210006), Hong Kong Research Grant Council (Grant Nos. 27209621, 17205922, 17212923), the National Natural Science Foundation of China under Grant 62204111. \textit{(Wei Xuan and Zihao Xuan contributed equally to this work.)} \textit{(Corresponding authors: Zhongrui Wang; Lang Feng.)}

Wei Xuan, Zihao Xuan, Zikang Yuan, Yuzhong Jiao and Luhong Liang are with the ACCESS – AI Chip Center for Emerging Smart Systems, InnoHK Centers, Hong Kong Science Park and the Hong Kong University of Science and Technology, Hong Kong, China (e-mail: \{weixuan, zihaoxuan, zikangyuan, yuzhongjiao, luhong\}@ust.hk).

Rongliang Fu and Tsung-Yi Ho are with The Chinese University of Hong Kong, Hong Kong 999077, China (e-mail: \{rlfu, tyho\}@cse.cuhk.edu.hk).

Ning Lin and Kwunhang Wong are with the University of Hong Kong, Hong Kong, China (e-mail: linning@hku.hk; u3556440@connect.hku.hk).

Zhongrui Wang is with Southern University of Science and Technology, Shenzhen, Guangdong, China (e-mail: wangzr@sustech.edu.cn).

Lang Feng is with Sun Yat-sen University Shenzhen Campus, Shenzhen, Guangdong, China (e-mail: flang1994@outlook.com).
}
}
\IEEEaftertitletext{\vspace{-2\baselineskip}}

\maketitle

\begin{abstract}

The rapid deployment of deep neural network (DNN) accelerators in safety-critical domains such as autonomous vehicles, healthcare systems, and financial infrastructure necessitates robust mechanisms to safeguard data confidentiality and computational integrity. Existing security solutions for DNN accelerators, however, suffer from excessive hardware resource demands and frequent off-chip memory access overheads, which degrade performance and scalability. 

To address these challenges, this paper presents a secure and efficient memory protection framework for DNN accelerators with minimal overhead. First, we propose a bandwidth-aware cryptographic scheme that adapts encryption granularity based on memory traffic patterns, striking a balance between security and resource efficiency. Second, we observe that both the overlapping regions in the intra-layer tiling's sliding window pattern and those resulting from inter-layer tiling strategy discrepancies introduce substantial redundant memory accesses and repeated computational overhead in cryptography. Third, we introduce a multi-level authentication mechanism that effectively eliminates unnecessary off-chip memory accesses, enhancing performance and energy efficiency. Experimental results show that this work decreases performance overhead by over $12\%$ and achieves $87\%$ energy efficiency improvement for both server and edge neural processing units (NPUs), while ensuring robust scalability.

\end{abstract}

\begin{IEEEkeywords}
Memory protection, authentication encryption, deep neural networks (DNNs), secure architectures
\end{IEEEkeywords}
\section{Introduction} \label{section:introduction}

\begin{figure}
    \centering
    \includegraphics[width=0.85\linewidth]{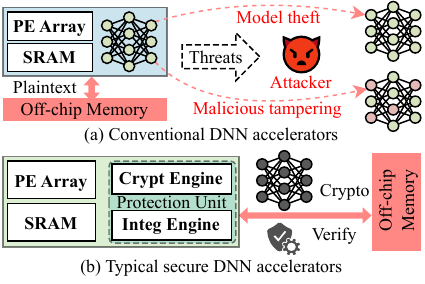}
    \caption{Overview of conventional and secure DNN accelerators. (a) Conventional DNN accelerators rely on untrusted off-chip memory and communication buses, rendering them vulnerable to model theft and malicious tampering. (b) Secure DNN accelerators protect data confidentiality and integrity through memory protection schemes employing authenticated encryption.}
    \label{figures:intro_overview}
\end{figure}

\IEEEPARstart{D}{eep} neural networks (DNNs)~\cite{krizhevsky2012imagenet,lecun2015deep,he2016deep} have revolutionized mission-critical applications across a wide range of domains, powering real-time decision-making in autonomous driving~\cite{grigorescu2020survey,parekh2022review}, healthcare~\cite{rong2020artificial,isensee2021nnu}, drug discovery~\cite{jumper2021highly}, chip design~\cite{mirhoseini2021graph}, smart manufacturing~\cite{wang2018deep,kotsiopoulos2021machine}, and finance~\cite{rundo2019machine}. The global artificial intelligence (AI) hardware market was valued at USD 54.10 billion in 2023 and is expected to surge to USD 474.10 billion by 2030, reflecting a remarkable CAGR of $38.73\%$~\cite{aihwmarket}. However, the widespread adoption of DNN-based AI systems has escalated concerns about insidious security threats for domain-specific accelerators that undermine their reliability, especially in high-stakes decision-making domains. For example, attackers can inject malicious data during the training phase to degrade model performance and accuracy, or introduce carefully crafted perturbations to input samples during inference to alter results.

Protecting DNN accelerators by ensuring confidentiality and integrity is paramount in modern AI systems for several reasons~\cite{lee2025secure}. \textit{Data Confidentiality}: The confidentiality of training datasets (often containing proprietary, privacy-sensitive, or mission-critical information) is non-negotiable. Protecting training data is essential to prevent unauthorized access and exploitation of private or sensitive information. Ensuring confidentiality helps maintain trust and compliance with data protection regulations. \textit{Resource Costliness}: The substantial investment in training resources necessitates stringent security protocols. For example, domain-specific hardware accelerators require substantial energy and financial investment, as the state-of-the-art model GPT-4~\cite{achiam2023gpt} requires tens of thousands of GPU-hours and millions to train. Additionally, iterative model tuning, hyperparameter optimization, and dataset cleaning necessitate significant human expertise and operational resources, which are irreplaceable if compromised. Safeguarding these resources not only ensures their optimal utilization but also protects against financial losses and inefficiencies. \textit{Vulnerability to Malicious Attacks}: DNN models are susceptible to various forms of malicious attacks, which can undermine their performance and reliability. Malicious intent poses a significant threat to DNN models. Protecting these models from potential tampering and other forms of interference is critical to maintaining their integrity and functionality. This protection is essential to prevent adverse outcomes that could arise from compromised models, thereby ensuring the overall reliability and trustworthiness of AI systems.

As shown in Fig.~\ref{figures:intro_overview}(a), various approaches~\cite{costan2016intel, zuo2021sealing, hua2022guardnn, hua2022mgx, lee2022tnpu, lee2023secureloop, shrivastava2023securator} have been proposed to secure conventional DNN accelerators against model theft and malicious tampering, mainly aiming to reduce off-chip memory access overhead for security metadata. They typically use the counter-mode encryption of Advanced Encryption Standard (AES-CTR) for confidentiality and the message authentication code (MAC) for integrity verification, as shown in Fig.~\ref{figures:intro_overview}(b). The counter value in AES-CTR concatenates the physical address (PA) and the version number (VN) of a data block, with the VN stored off-chip and incremented with each write. To ensure the integrity of off-chip memory data, each data block is accompanied by a message authentication code (MAC). Additionally, a Merkle Tree (MT)~\cite{gassend2003caches} and its variants are often utilized, with the root stored on-chip to prevent replay attacks.

In this paper, we propose a novel secure and efficient DNN accelerator framework, which achieves reduced hardware resource overhead and negligible performance overhead for integrity verification, while maintaining the same level of security and practical applicability. This paper makes the following contributions:

\begin{itemize}[leftmargin=*]
    \item {\textbf{Insight.} Providing an in-depth insight into the limitations of current memory protection strategies for DNN accelerators highlights two critical concerns: the substantial \textit{hardware overhead for encryption} and the expensive \textit{off-chip memory access for integrity verification}.
    }
    \item {\textbf{Solution.} Through hardware/software synergy, we present a secure and efficient accelerator framework. It incorporates a \textit{bandwidth-aware cryptographic scheme} that utilizes a single AES engine with adjustable encryption granularity to minimize hardware resource overhead. Furthermore, its \textit{multi-level authentication mechanism} significantly reduces or eliminates off-chip memory access overhead.
    }
    \item {\textbf{Evaluation.} Conducting comprehensive experiments using cycle-accurate simulators for DNN accelerators, memory protection schemes, and off-chip memory accesses. Experimental results demonstrate that our proposed framework achieves $12.3\%$ performance speedup and $87.6\%$ energy efficiency improvement compared to state-of-the-art methods, while maintaining robust scalability with minimal hardware overhead to satisfy the bandwidth demands of accelerators.
    }
\end{itemize}

This paper is structured as follows: Section~\ref{section:preliminaries} reviews related work on memory protection and secure DNN accelerators. Section~\ref{section:threat_moti} introduces our threat model and motivation. In Section~\ref{section:archi}, we introduce our proposed framework. We then conduct extensive experiments on various DNN models in Section~\ref{section:evaluation}. Finally, Section~\ref{section:conclusion} concludes our work.
\lstdefinelanguage[riscv]{Assembler}%
{
morestring=[b]",
morekeywords = [1]{.alias,.align,.ascii,.asciiz,.byte,.data,.double,.end,.endb,%
  .endr,.ent,.err,.extern,.file,.float,.fmask,.frame,.globl,.half,.text,%
  .verstamp,.vreg,.word},%
morekeywords = [2]{lw,sw,lb,lbu, lh, lhu, sb, sh, la,li,lui, ori, or, sub, subi, add, addi, or, ori, xor, xori, sra, srai, srl, srli, slli, srl, sll, beq, bne,bge,blt, bltu, bgeu,j,jal,jalr, jr, j, or,addi,ecall,ret,load_patch, call, push, pop, jmp},%
morekeywords = [3]{x0,x1,x2,x3,x4,x5,x6,x7,x8,x9,x10,x11,x12,x13,x14,x15,x16,x17,x18,x19,x20,x21,ra,s0,s1,s2,a0,a1,x28,x29,x30,x31,sp,zero, a0,a1,a2,a3,a4,a5,a6,a7,sp,ra,t0,t1,t2,t3,t4,t5,t6,t7,s0,s1,s2,s3,s4,s5,s6,s7,s11},%
comment = [l]\#,%
keywordsprefix=.,%
sensitive=false,%
}[keywords,comments,strings]

\lstdefinestyle{customriscv}{
  language = [riscv]Assembler,
  basicstyle = \footnotesize\ttfamily,
  stringstyle = \footnotesize\ttfamily,
  frame = single,
  rulesep = 5pt,
  numbers = none,
  framexleftmargin = \parindent,
  showspaces = false,
  showstringspaces = false,
  showtabs = false,
  rulecolor = \color{gray},
  rulesepcolor = \color{gray},
  tabsize=3,
  breaklines=false,
  keywordstyle = [1]\color{purple},%
  keywordstyle = [2]\color{blue},%
  keywordstyle = [3]\color{red},%
  stringstyle = \color{mauve},%
  escapeinside = {\%*}{*)},
  moredelim=[is][\bfseries\color{violet}]{;?}{?;},
  moredelim=[is][\bfseries\color{red}]{;!}{!;},
  moredelim=[is][\bfseries\color{green}]{;@}{@;},
  moredelim=[is][\bfseries]{[*}{*]},
  literate={à}{{\`a\ }}1 {é}{{\'e}}1 {è}{{\`e}}1 {ê}{{\^e}}1, %
  extendedchars=true
}

\lstset{style=customriscv}
\section{Preliminaries} \label{section:preliminaries}

\subsection{Convolutional Computation} \label{cover:prelims_cc}

\begin{figure}[t]
    \centering
    \includegraphics[width=0.8\linewidth]{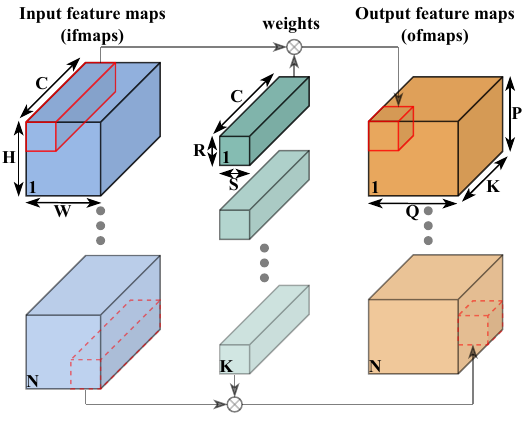}
    \caption{CNN convolution operation (CONV).}
    \label{figures:prelims_cnn}
\end{figure}

Convolutional neural networks (CNNs) are a dominant architecture in DNNs, where convolutional (CONV) layers account for over $90\%$ of operations and substantial data movement~\cite{krizhevsky2012imagenet,simonyan2014very,chen2016eyeriss,ma2018optimizing}, as shown in Fig.~\ref{figures:prelims_cnn}. The CONV layer efficiently extracts features from high-dimensional data through local connectivity and weight-sharing mechanisms, employing learnable convolutional kernels that slide over input data to perform localized dot product operations and generate feature maps that reflect the activation strength of specific patterns. Computationally, the convolutional operation can be expressed as a 7-dimensional nested loop (Fig.~\ref{listings:prelims_cnn_loop}), processing input feature maps (ifmaps) with dimensions (N, C, H, W) and filter weights with dimensions (K, C, R, S) to generate output feature maps (ofmaps) with dimensions (N, K, P, Q), where N represents batch size, H/W and P/Q denote spatial dimensions of input and output respectively, C/K represent input/output channels, and R/S specify filter spatial dimensions.

\begin{figure}[t]
    \centering
\begin{lstlisting}[language=Python]
# Variables: N(batches); K(output channels);
# C(input channels); R,S: kernel dimensions;
# P,Q: output dimensions; H,W: input dimensions; 
for n,k,c,p,q,r,s in range(N,K,C,P,Q,R,S):
    h = p * stride - padding + r
    w = q * stride - padding + s
    ofmaps[n][k][p][q] += ifmaps[n][c][h][w] * 
      weights[k][c][r][s]
\end{lstlisting}
    \vspace{-5pt}
    \caption{Example of 7-dimensional nested loop for CONV.}
    \label{listings:prelims_cnn_loop}
    \vspace{-10pt}
\end{figure}

\subsection{Confidentiality Protection} \label{cover:prelims_cp}

\begin{figure}[t]
    \centering
    \includegraphics[width=0.8\linewidth]{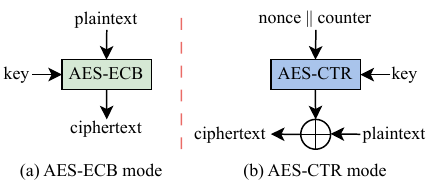}
    \caption{Comparison of AES encryption modes: (a) AES-ECB mode uses only a global encryption key; (b) AES-CTR mode uses a unique value (nonce $||$ counter) and a global key.}
    \label{figures:prelims_aes_ecb_ctr}
    \vspace{-10pt}
\end{figure}

Data encryption/decryption, notably the extensively researched Advanced Encryption Standard (AES)~\cite{daemen2002design}, is a dominant security countermeasure against data confidentiality attacks~\cite{suh2003caches,suh2003efficient,rogers2007using,zubair2019anubis,zuo2019supermem}. The electronic codebook mode (AES-ECB) provides a direct approach for encrypting/decrypting data blocks, as illustrated in Fig.~\ref{figures:prelims_aes_ecb_ctr}(a). Before data enters the processor chip, AES-ECB first decrypts blocks read from off-chip memory, and the resulting plaintext is then processed within the secure chip. Conversely, when writing to off-chip memory, data blocks are first encrypted, and the generated ciphertext is subsequently written. However, AES-ECB presents significant limitations: its inadequate diffusion properties cause identical plaintext blocks to consistently produce identical ciphertext blocks, thereby exposing underlying data patterns~\cite{schneier2007applied,menezes2018handbook}. Moreover, its inherently sequential processing model imposes substantial performance penalties, and memory reads in particular form a critical bottleneck during program execution.

To address these limitations, confidentiality protection mechanisms widely adopt counter mode encryption (AES-CTR), which enhances security through unique counters to resist repeating pattern attacks and reduces latency via the XOR-based one-time pad (OTP) technique~\cite{lipmaa2000ctr,shi2005high,yan2006improving,liu2018crash,zuo2019supermem}. As shown in Fig.~\ref{figures:prelims_aes_ecb_ctr}(b), AES-CTR uses a non-repeating and incrementing counter to produce an OTP for each encryption/decryption under the same key. The counter concatenates the physical address of a data block and the version number that is incremented on each write memory operation. Let $K_e, \mathcal{P}, \mathcal{C}$ denote the AES key, plaintext, and ciphertext, respectively. The AES-CTR mode for encryption and decryption can be formulated as Eq.~\ref{equs:aes_ctr_enc} and Eq.~\ref{equs:aes_ctr_dec}, where $\text{AES}\mbox{-}\text{CTR}_{K_e}(\text{PA} || \text{VN})$ produces an OTP, with $||$ and $\oplus$ representing bit-wise concatenation and XOR operators, respectively.
\begin{align}
    \mathcal{C} &= \mathcal{P} \oplus \text{AES}\mbox{-}\text{CTR}_{K_e}(\text{PA} \; || \; \text{VN}) \label{equs:aes_ctr_enc} \\
    \mathcal{P} &= \mathcal{C} \, \oplus \text{AES}\mbox{-}\text{CTR}_{K_e}(\text{PA} \; || \; \text{VN}) \label{equs:aes_ctr_dec}
\end{align}

\begin{figure}[t]
    \centering
    \includegraphics[width=0.98\linewidth]{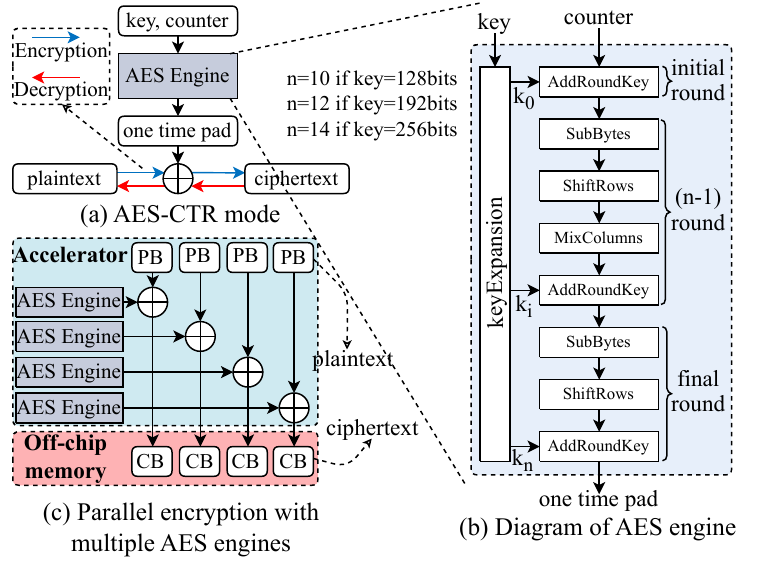}
    \caption{Summary of AES-CTR mode. (a) Reusing the AES engine for encryption and decryption in AES-CTR mode. (b) Diagram of the AES engine, featuring the AddRoundKey, SubBytes, ShiftRows, MixColumns, and KeyExpansion modules. (c) Utilization of multiple AES engines for parallel encryption to boost high-bandwidth capabilities.}
    \label{figures:prelims_aes_ctr_mode}
    \vspace{-10pt}
\end{figure}

Rather than directly applying AES-ECB to data blocks, AES-CTR generates an OTP (one-time pad) using a dynamically unique nonce-counter pair and performs encryption/decryption through a simple bitwise XOR with this pad. Thus, AES-CTR enables parallelization of memory reads with decryption or computation with encryption, thereby allowing the overlap of additional latency introduced by confidentiality protection. When an accelerator reads a data block from off-chip memory, the protection module uses the version number (VN) to generate an OTP and decrypts the block by XORing it with the OTP (red path in Fig.~\ref{figures:prelims_aes_ctr_mode}(a), Eq.~\ref{equs:aes_ctr_dec}). Similarly, when writing a data block to off-chip memory, the module increments the VN to generate an OTP and encrypts the block by XORing it with the OTP (blue path in Fig.~\ref{figures:prelims_aes_ctr_mode}(a), Eq.~\ref{equs:aes_ctr_enc}). Fig.~\ref{figures:prelims_aes_ctr_mode}(b) illustrates the diagram of the AES engine along with its key functional modules.

\subsection{Integrity Verification}

To safeguard data integrity against potential off-chip memory tampering in secure computing systems, a cryptographically robust message authentication code (MAC) mechanism is typically employed. During write operations to untrusted off-chip memory, the system generates a unique MAC through a cryptographic hash function that incorporates both the data payload and a secret authentication key. When retrieving data from off-chip storage, the system recalculates the MAC using the stored data and verifies its cryptographic consistency with the originally stored MAC value. This approach ensures that any unauthorized modifications to the data can be detected through MAC mismatch, providing fundamental data authenticity guarantees for secure computing environments.

However, relying solely on MAC for individual data blocks fails to guarantee temporal freshness and opens vulnerabilities to replay attacks, where adversaries can substitute current data with previously valid but outdated versions. To counter this sophisticated threat, prior works utilize hierarchical integrity trees for comprehensive MAC verification, such as Merkle Tree (MT)~\cite{gassend2003caches,merkle2019protocols} and Bonsai Merkle Tree (BMT)~\cite{rogers2007using}, with the tree root securely stored on-chip to prevent malicious tampering. These tree structures create cryptographic dependencies between data blocks and their associated metadata, ensuring that any replay attempt will result in inconsistent tree traversal. The overhead of integrity verification is non-trivial since it requires traversing both MACs and multiple levels of tree nodes stored in off-chip memory, introducing additional memory accesses and computational complexity to prevent replay attacks.

\subsection{Prior Secure DNN Accelerators} \label{sec:prem:prior_secure_dnn_acc}

Trusted Execution Environments (TEEs) have been well-established for general-purpose processors. Specifically, Intel SGX~\cite{costan2016intel} creates secure enclaves using CPU hardware mechanisms with AES encryption and Merkle Trees (MTs) rooted in the Trusted Computing Base (TCB). ARM TrustZone~\cite{pinto2019demystifying} partitions processors into secure and normal worlds through isolated execution environments, enabling trusted applications to run in a protected domain. Extending these paradigms to specialized accelerators, SEAL~\cite{zuo2021sealing} introduces criticality-aware smart encryption that selectively bypasses encryption for partial data while colocating data with counters. GuardNN~\cite{hua2022guardnn} achieves privacy-preserving deep learning through minimal TCB design. However, these approaches still incur substantial memory traffic overhead due to reliance on MTs or integrity trees.

Recent works have focused on specialized optimizations to mitigate security metadata overhead in DNN accelerators. For integrity tree optimization, \cite{feng2023efficient} proposes Migratable Merkle Tree (MMT) with integrity forests for distributed memory, while \cite{xu2024data} introduces TEE-oriented data abstraction with flexible permission controls for efficient inter-enclave collaboration. For VN management, MGX~\cite{hua2022mgx} employs on-chip status tracking for VN generation with coarse-grained integrity verification, TNPU~\cite{lee2022tnpu} generates all VNs within on-chip caches to eliminate external dependencies, and SoftVN~\cite{umar2022softvn} enables software-provided VNs to avoid off-chip VN accesses. For specialized mechanisms, TensorTEE~\cite{han2024tensortee} proposes unified tensor-granularity for secure collaborative computing, Securator~\cite{shrivastava2023securator} employs layer-level integrity checking by XORing MACs within layers, and SecureLoop~\cite{lee2023secureloop} dynamically searches for optimal scheduling strategies. Despite these advances, existing solutions address only isolated aspects of security overhead (e.g., reducing off-chip metadata accesses, optimizing redundant authorization for overlapped data, or mitigating security-related bandwidth loss) and lack holistic optimization across all critical dimensions. Consequently, they achieve limited overall performance improvements. In contrast, our solution simultaneously addresses multiple overhead sources through a bandwidth-aware cryptographic scheme and multi-level authentication mechanism, achieving comprehensive performance optimization.
\section{Threat Model and Motivation} \label{section:threat_moti}

\subsection{Threat Model} \label{sec:threat_moti:threat_model}

Consistent with established threat models in secure computing literature~\cite{gassend2003caches,hua2022mgx, lee2022tnpu, xuan2025seda}, we adopt the trusted computing base paradigm where the accelerator hardware itself is presumed secure, ensuring that its internal computational state and on-chip resources remain inaccessible to adversaries through direct observation or manipulation. This security perimeter encompasses all on-chip components, including processing units, control logic, and internal memory structures, which are protected by hardware-based isolation mechanisms and tamper-resistant design principles.

However, all external components interfacing with the accelerator are classified as untrusted entities, encompassing off-chip memory subsystems, interconnection buses, and peripheral communication channels. Within this adversarial model, attackers are assumed to possess comprehensive capabilities to monitor, intercept, modify, or replay any data traversing these untrusted channels through sophisticated physical intrusion techniques or malicious software exploitation. Notably, our threat model excludes advanced physical side-channel attacks designed to infer DNN architectural details~\cite{yan2020cache}, such as power consumption analysis and electromagnetic emission monitoring. Furthermore, algorithmic adversarial attacks~\cite{akhtar2018threat} targeting machine learning model vulnerabilities fall outside our security scope, as our focus centers on cryptographic access control mechanisms that restrict data decryption to authorized entities possessing valid cryptographic credentials.

\subsection{Motivation} \label{sec:threat_moti:motivation}

\begin{figure}[t]
  \centering
  \subfloat[Memory access overhead]{\includegraphics[height = 0.37\linewidth]{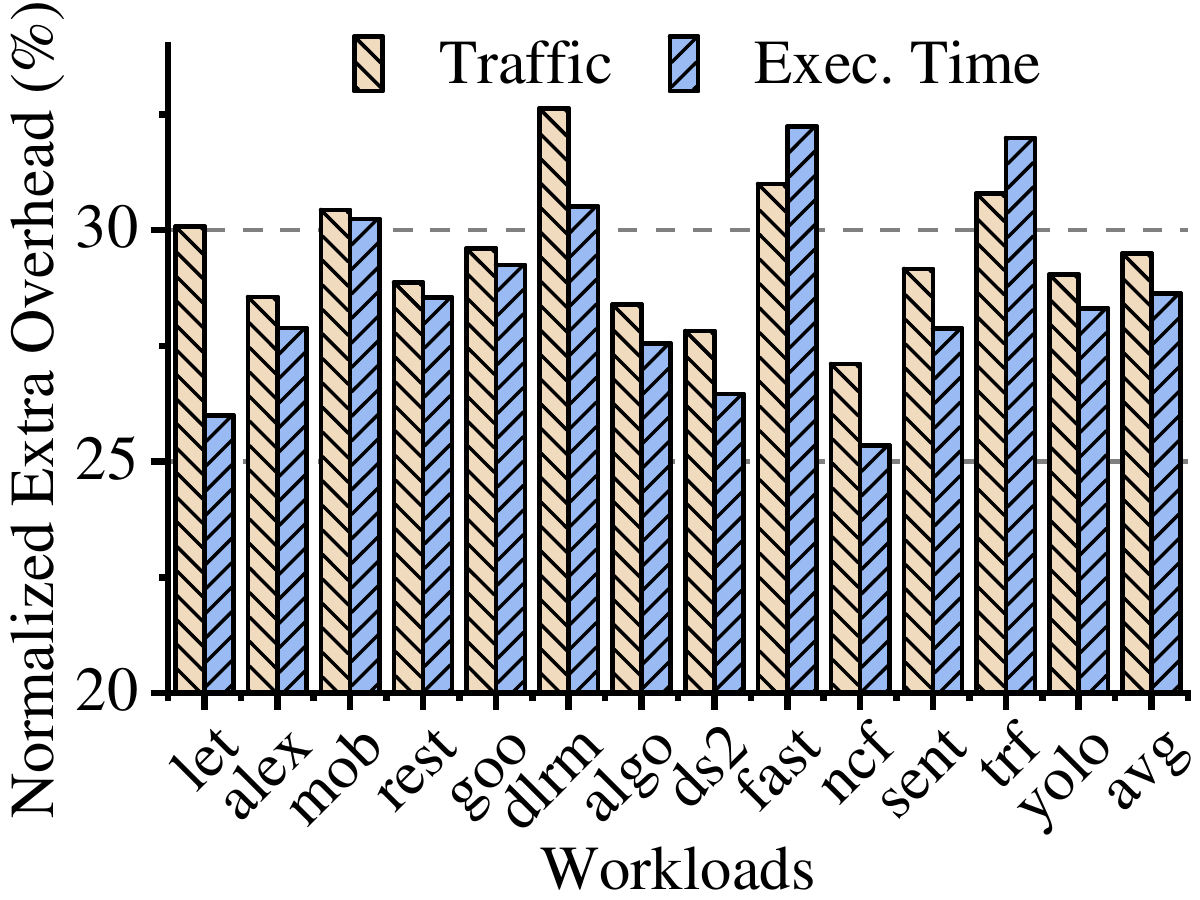}}
  \hfill
  \subfloat[Hardware crypto overhead]{\includegraphics[height = 0.38\linewidth]{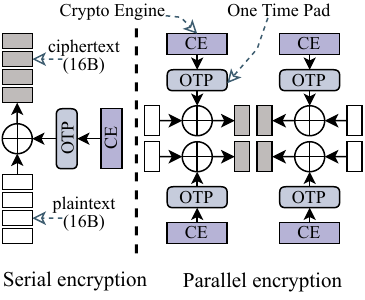}}
  \caption{Motivations of our work: (a) Accessing security metadata in off-chip memory introduces additional latency and memory traffic. (b) Serial encryption with a single crypto engine provides insufficient throughput, while parallel encryption with multiple engines incurs excessive hardware overhead.}
  \label{figures:threat_motivation}
  \vspace{-10pt}
\end{figure}

\subsubsection{Memory Access Overhead Challenge for Integrity Verification} \label{sec:threat_moti:motivation:mem_access}
Ensuring the confidentiality and integrity of untrusted off-chip memory through security metadata incurs substantial memory access overhead. As shown in Fig.~\ref{figures:threat_motivation}(a), accessing authentication metadata introduces additional memory traffic and execution time penalties across diverse DNN workloads. While existing approaches propose optimizations, including Bonsai Merkle Tree (BMT)~\cite{rogers2007using} for reduced storage and faster verification, coarser-grained protection units (e.g., 512B blocks versus 64B cachelines), and dynamic version number (VN) updates based on DNN model state~\cite{hua2022mgx,lee2022tnpu}, these methods fail to fully address DNN-specific challenges. For instance, Securator~\cite{shrivastava2023securator} employs layer-level MACs to reduce off-chip memory access for integrity verification. However, it overlooks intra-layer tile overlaps and inter-layer tiling pattern variations, potentially causing redundant encryption/decryption operations, excessive integrity verification overhead, and false negatives when tiling patterns differ across layers.

\subsubsection{Encryption Bandwidth Challenge for Confidentiality Protection} \label{sec:threat_moti:motivation:enc_bandwidth}
AES engines can only encrypt 128-bit blocks per operation, creating a fundamental bandwidth-hardware tradeoff for DNN accelerators. As shown in Fig.~\ref{figures:threat_motivation}(b), serial encryption with a single crypto engine provides insufficient throughput to match accelerator bandwidth demands, while parallel encryption with multiple engines achieves adequate throughput at the cost of prohibitive hardware resource consumption. For example, Securator~\cite{shrivastava2023securator} deploys four parallel AES engines to encrypt/decrypt 64-byte data blocks, as shown in Fig.~\ref{figures:prelims_aes_ctr_mode}(c). This approach imposes significant strain on resource-constrained accelerators, necessitating a careful balance between hardware resource allocation and cryptographic security performance. Neither insufficient serial throughput nor excessive parallel hardware overhead achieves the optimal bandwidth-efficiency balance required for practical secure DNN accelerator deployment.

\section{Secure Framework} \label{section:archi}

\begin{figure*}
    \centering
    \includegraphics[width=0.98\linewidth]{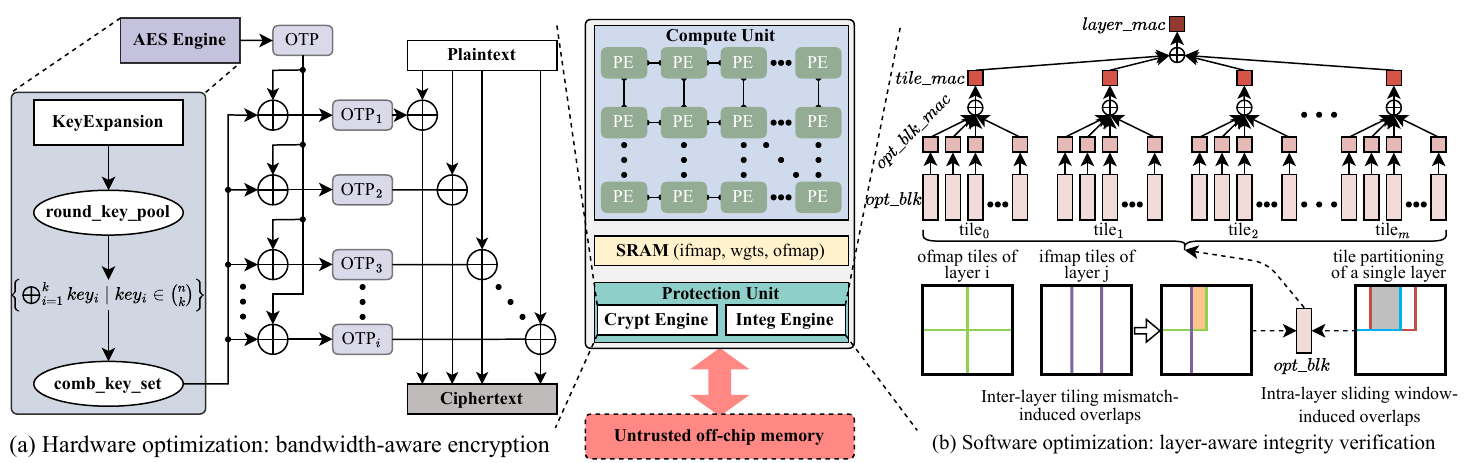}
    \caption{Overview of proposed architecture. (a) The Crypt Engine optimizes hardware by XORing keys from the AES Engine's KeyExpansion module with OPT, creating bandwidth-sensitive encryption granularity and reducing hardware overhead compared to using multiple AES Engines. (b) The Integ Engine optimizes software by analyzing overlapping tiles within a layer and patterns across layers to determine the optimal block($opt\_blk$) for integrity verification. This leads to a multi-level integrity verification mechanism with $opt\_blk\_mac$, $tile\_mac$, and $layer\_mac$.}
    \label{figures:arch_framework}
    \vspace{-10pt}
\end{figure*}

\subsection{Overview of Secure DNN Accelerator Framework} \label{sec:arch:overview}

Fig.~\ref{figures:arch_framework} presents our comprehensive hardware-software co-design framework for secure DNN accelerators. The framework addresses two fundamental yet interdependent security challenges through synergistic optimization. On the hardware front (Fig.~\ref{figures:arch_framework}(a)), we tackle the bandwidth-scalability dilemma of cryptographic engines, where conventional approaches that replicate AES engines to meet accelerator bandwidth demands incur prohibitive hardware costs. Our bandwidth-aware cryptographic scheme resolves this challenge by augmenting a single AES engine with minimal XOR logic gates, leveraging the KeyExpansion module to generate multiple unique OTPs without proportional hardware scaling. On the software front (Fig.~\ref{figures:arch_framework}(b)), we exploit the deterministic execution patterns inherent to DNNs. By statically analyzing intra-layer and inter-layer dependencies across the entire model, our layer-aware integrity verification mechanism determines optimal authentication block sizes ($opt\_blk$) that eliminate redundant verification overhead for overlapping tiles while establishing a multi-level MAC hierarchy ($opt\_blk\_mac$, $tile\_mac$, $layer\_mac$). The synergy between these hardware and software optimizations proves essential, as hardware efficiency enables high encryption bandwidth without resource explosion, while software minimizes integrity verification traffic.

The primary hardware challenge lies in cryptographic engine scalability. AES engines encrypt only 128-bit blocks per operation, creating bandwidth bottlenecks for high-throughput DNN accelerators. Conventional approaches deploy multiple parallel AES engines—Securator~\cite{shrivastava2023securator} employs four parallel AES-128 engines for 64-byte blocks. However, this linear scaling imposes severe resource penalties on area-constrained accelerators, particularly edge devices with limited hardware budgets. Our bandwidth-aware cryptographic scheme (Fig.~\ref{figures:arch_framework}(a)) decouples bandwidth scaling from proportional hardware growth through architectural innovation. The primary software challenge concerns security metadata overhead. Even with compact encoding, an 8-byte MAC protecting a 64-byte block increases memory traffic by 12.5\% ($\frac{8}{64}$). This overhead compounds with inter-layer dependencies, where misaligned authentication blocks and tiling patterns create cascading inefficiencies through redundant metadata propagation and re-authentication of shared data. Our multi-level authentication mechanism (Fig.~\ref{figures:arch_framework}(b)) exploits DNN computational determinism by storing aggregated MACs at layer or model granularity within on-chip SRAM, substantially reducing or eliminating off-chip metadata traffic while preserving cryptographic security.

\subsection{Bandwidth-Aware Cryptographic Scheme} \label{sec:arch:bandwidth_aware}

Currently, most DNN accelerators employing AES mode fix their encryption granularity at 128 bits. To meet the high bandwidth demands of accelerators, an expansion in the quantity of hardware encryption engines for AES is imperative. This leads to heightened consumption of hardware resources and power for ensuring confidentiality, especially challenging for edge devices with limited hardware capacities. Additionally, using the same OTP for all data in a block can pose security risks. This dilemma often results in a suboptimal trade-off between security and computational efficiency. To mitigate this challenge, we leverage the inherent characteristics of AES to achieve bandwidth-aware encryption granularity through a single AES-CTR encryption engine. This approach balances hardware consumption with effective bandwidth imposed by confidentiality protection.

\subsubsection{Exploration of the Trade-off Between Hardware Resource Consumption and Encryption/Decryption Bandwidth} Table~\ref{table:arch_aes_spec} presents the specifications of various hardware implementations of AES~\cite{banerjee2017energy, lee2023secureloop}. Among them, the pipelined AES offers the highest bandwidth of 16 B/cycle, but its area is $26.27 \times$ larger than that of serial AES. When considering both performance and area, parallel AES emerges as the preferred choice, as it provides relatively high bandwidth without incurring significant area overhead.

While maintaining an equivalent level of security, numerous approaches have bolstered encryption parallelism by stacking multiple AES engines. For instance, Securator~\cite{shrivastava2023securator} uses four AES engines to concurrently encrypt a $64$B data block, as shown in Fig.~\ref{figures:prelims_aes_ctr_mode}(c). SecureLoop~\cite{lee2023secureloop} uses three AES-GCM engine implementations tailored for different data types, including ifmaps, weights, and ofmaps. Nevertheless, the drawback of sacrificing hardware resources to achieve performance improvements in encryption and decryption is apparent, particularly for edge DNN accelerators constrained by limited hardware capabilities. This situation emphasizes the crucial equilibrium needed to optimize security and operational efficiency in these systems.

\begin{table}[t]
    \caption{Specifications for multiple AES hardware implementations.}
    \label{table:arch_aes_spec}
    \centering
    \footnotesize
    \renewcommand{\arraystretch}{0.85}
    \setlength{\tabcolsep}{1.8pt}
    \begin{tabular}{c|c|c|c}
        \toprule
        \multicolumn{1}{c|}{\textbf{AES implementations}}   & \multicolumn{1}{c|}{\textbf{Performance (cycles)}} & \multicolumn{1}{c|}{\textbf{Area (gates)}} & \multicolumn{1}{c}{\textbf{Energy (pJ)}} \\ 
        \midrule                   
        \textbf{Pipelined}   &  $1$  &  $78.8$k  & $165.1$ \\
        \textbf{Parallel}    &  $11$  & $9.2$k &  $194.6$ \\
        \textbf{Serial}      &  $336$ & $3.0$k &  $768.0$ \\
        \bottomrule
    \end{tabular}
\end{table}

\subsubsection{Security Threat Assessment of Shared One-Time Pad} \label{sec:arch:bandwidth:sec_threat_access}

\begin{algorithm}[t]
	\DontPrintSemicolon
	\SetKwInOut{Input}{\textbf{Input}}	
	\SetKwInOut{Output}{\textbf{Output}}	
	\SetKwFunction{Attack}{\textbf{Attack of SECA}}{}{}
	\SetKwFunction{Defense}{\textbf{Defense of SECA}}{}{}
    \let\oldnl\nl
	\newcommand{\nonl}{\renewcommand{\nl}{\let\nl\oldnl}}

	\Input{OTP: one time pad for a data block ($blk$); $most\_value\_p$: most used plaintext in $blk$.}
	\Output{$value\_p$: all plaintext of $blk$ \\}

    $most\_value\_p \leftarrowtail $ \textsc{StatAnalysis}($blk$) \label{alg:arch_seca_attack_stat} \\
    $most\_value\_c \leftarrowtail $ \textsc{CalcFreqValue}($blk$) \label{alg:arch_seca_attack_calc} \\
    OTP$ \leftarrowtail most\_value\_p \oplus most\_value\_c$ \label{alg:arch_seca_attack_otp} \\
    \For{each encrypted element value\_c of blk} { \label{alg:arch_seca_attack_for:begin}
        $value\_p \leftarrowtail value\_c \; \oplus$ OTP \label{alg:arch_seca_attack_for_end} \\
    }
	\caption{Security threat of SECA} \label{alg:arch_seca_attack}
\end{algorithm}

\begin{figure}
    \centering
    \includegraphics[width=0.8\linewidth]{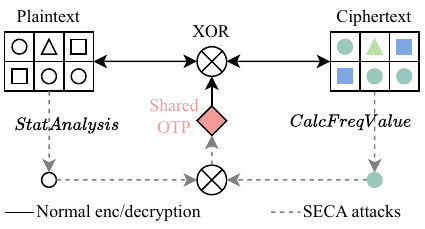}
    \caption{Example of SECA attacks. The data block consists of six 128-bit sub-blocks. By utilizing \textsc{StatAnalysis} and \textsc{CalcFreqValue}, attackers can obtain the corresponding plaintext and ciphertext for the most frequent sub-block. Then by XORing these two values, attackers derive the shared OTP, and decrypt all ciphertext.}
    \label{figures:arch_seca_example}
\end{figure}

A straightforward approach is to use each engine once per data block, with each 128-bit segment within this data block sharing the same OTP. Nevertheless, assigning an OTP to an individual data block, particularly containing multiple 128-bit sub-blocks, poses security risks and could be potentially vulnerable to a \underline{S}ingle-\underline{E}lement \underline{C}ollision \underline{A}ttack (\text{SECA}).

The principle of SECA, which exploits scenarios where multiple 128-bit sub-blocks share the same OTP, is outlined in lines~\ref{alg:arch_seca_attack_stat}-\ref{alg:arch_seca_attack_for_end} of Algorithm~\ref{alg:arch_seca_attack}. Under our threat model in Section~\ref{sec:threat_moti:threat_model}, an attacker can observe the ciphertext of all 128-bit sub-blocks. Furthermore, the attacker employs the \textsc{CalcFreqValue} function to identify the most frequently occurring encrypted value within the block ($most\_value\_c$). By exploiting the known sparsity characteristics inherent to DNN data distributions, the attacker can probabilistically infer the most common plaintext value $most\_value\_p$ (typically $most\_value\_p=0$).

Referring to the AES-CTR encryption formula Eq.~\ref{equs:aes_ctr_enc}, by computing the XOR of these two values ($most\_value\_c$ and $most\_value\_p$), the shared OTP for the block can be reverse-engineered, as shown in line~\ref{alg:arch_seca_attack_otp} of Algorithm~\ref{alg:arch_seca_attack}. Since the block shares this OTP, once obtained, we can derive all plaintext values within the encrypted block, as shown in lines~\ref{alg:arch_seca_attack_for:begin}-\ref{alg:arch_seca_attack_for_end} of Algorithm~\ref{alg:arch_seca_attack}. Extending this principle further, we can extract the data values from every DNN layer and even the entire model. Fig.~\ref{figures:arch_seca_example} depicts a simplified conceptual example of an adversary launching the SECA.

\subsubsection{Proposed Approach} \label{sec:arch:bandwidth_aware:approach}

We introduce a novel bandwidth-aware cryptographic scheme to meet the high bandwidth requirements of accelerators without compromising the security of encryption. While re-keying serves as a critical cryptographic primitive for extending key lifetimes in long-term deployments~\cite{setia2000kronos, medwed2010fresh}, it addresses an orthogonal problem to our bandwidth-aware scheme. Unlike re-keying, our work resolves the core trade-off between encryption bandwidth demands and hardware resource constraints via cryptographic optimization, enabling efficient high-bandwidth encryption tailored explicitly for resource-constrained DNN accelerators. Our method leverages the inherent features of AES-CTR encryption mode, using just a single AES engine and a minimal number of XOR logic gates, as illustrated in Fig.~\ref{figures:arch_framework}(a).

\begin{table}[t]
\caption{Analysis of unique OTP combinations for different AES key-length variants.}
\label{table:arch_aes_variants}
\centering
    \footnotesize
    \renewcommand{\arraystretch}{0.95}
    \setlength{\tabcolsep}{1.8pt}
    \begin{tabular}{c|c|c|c}
    \toprule
        \multicolumn{1}{c|}{\textbf{AES variants}}   & \multicolumn{1}{c|}{\textbf{Initial key length}} & \multicolumn{1}{c|}{\textbf{$\#$Round key}} & \multicolumn{1}{c}{\textbf{Combination key set}} \\ 
    \midrule                  
        \textbf{AES-128} & $16$B & $10$ & $1024$ \\
        \textbf{AES-192} & $24$B & $12$ & $4096$ \\
        \textbf{AES-256} & $32$B & $14$ & $16384$ \\
    \bottomrule
    \end{tabular}
\end{table}

The Advanced Encryption Standard (AES) employs a key scheduling algorithm to expand a base key into a series of distinct round keys. This scheduling process generates the necessary round keys from the initial key. Depending on the length of the initial key, the three AES variants feature different numbers of encryption rounds: AES-128 uses a 16-byte initial key with 10 rounds, AES-192 utilizes a 24-byte initial key with 12 rounds, and AES-256 adopts a 32-byte initial key with 14 rounds, as illustrated in Table~\ref{table:arch_aes_variants}. Here, the ``Combination Key Set'' ($comb\_key\_set$) refers to the total combinations of all round keys. The ``Round Key'' corresponds to the number of rounds, which is equal to the total number of generated round keys (abbreviated as $round\_key\_set$). We can obtain $comb\_key\_set$ by the following equation: $comb\_key\_set=\sum_{i=0}^{round\_key\_set} C(round\_key\_set,i)=2^{round\_key\_set}$.

\begin{algorithm}[tb]
    \DontPrintSemicolon
    \SetKwInOut{Input}{\textbf{Input}}	
    \SetKwInOut{Output}{\textbf{Output}}
    \let\oldnl\nl
    \newcommand{\nonl}{\renewcommand{\nl}{\let\nl\oldnl}}

    \Input{PA, VN: physical address and version number of a data block; n: denotes the required bandwidth divided by the single AES engine bandwidth.}
    \Output{$block\_otp\_i$: unique OTP for each $128$-bit sub-block$_i$ within this data block.}

    $round\_key\_set \leftarrowtail$ \textsc{KeyExpansion}($initial\_key$) \label{alg:arch_seca_defense_keyexpan} \\
    $shared\_otp \leftarrowtail$ \textsc{AES-CTR}(PA $||$ VN, $round\_key\_set$) \label{alg:arch_seca_defense_aes_ctr} \\
    $comb\_key\_set \leftarrowtail \textsc{CalcComb}(round\_key\_set)$ \label{alg:arch_seca_defense_calc_comb} \\
    $comb\_key\_n \leftarrowtail \textsc{RandSelect}(comb\_key\_set, n)$ \label{alg:arch_seca_defense_rand_select} \\
    \For{each $128\mbox{-}$bit comb\_key\_i in comb\_key\_n} { \label{alg:arch_seca_defense_for_begin}			
        $block\_otp\_i \leftarrowtail (shared\_otp \oplus comb\_key\_i) $ \label{alg:arch_seca_defense_for_end} \\
    }

    \caption{Defense against SECA} \label{alg:arch_seca_defense}
\end{algorithm}

\begin{figure}[t]
    \centering
    \includegraphics[width=0.65\linewidth]{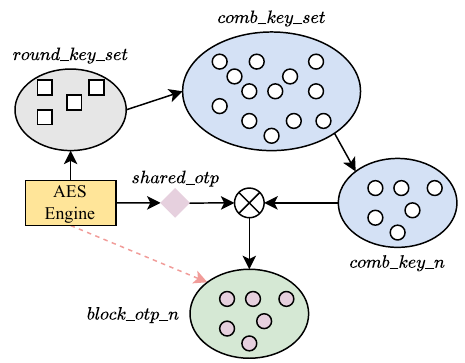}
    \caption{Description of generating unique block OTPs for multiple 128-bit sub-blocks within a data block.}
    \label{figures:arch_seca_otps}
    \vspace{-10pt}
\end{figure}

To defend against SECA attacks while meeting high bandwidth demands, we propose a bandwidth-aware cryptographic scheme with negligible hardware overhead. As detailed in Algorithm~\ref{alg:arch_seca_defense}, we leverage the KeyExpansion module and AES-CTR engine (\textsc{AES-CTR}) to obtain $round\_key\_set$ and $shared\_otp$ (lines~\ref{alg:arch_seca_defense_keyexpan}-\ref{alg:arch_seca_defense_aes_ctr}). Based on bandwidth requirements, \textsc{RandSelect} chooses $comb\_key\_n$ from the total round key combinations computed by \textsc{CalcComb} (line~\ref{alg:arch_seca_defense_calc_comb}). We then generate multiple unique OTPs ($block\_otp\_n$) by XORing $shared\_otp$ with selected combination keys $comb\_key\_n$ (lines~\ref{alg:arch_seca_defense_for_begin}-\ref{alg:arch_seca_defense_for_end}). This ensures each 128-bit sub-block receives a unique OTP ($block\_otp\_i$), effectively preventing SECA attacks. Critically, this approach conserves hardware resources by using minimal XOR logic gates instead of multiple parallel AES engines. Fig.~\ref{figures:arch_seca_otps} depicts the scenario of generating unique block OTPs ($block\_otp\_n$) for multiple 128-bit sub-blocks within a data block. Instead of using $n$ AES Engines to generate $n$ unique OTPs, we leverage the $round\_key\_set$ and $shared\_otp$ to produce the required $block\_otp\_n$ with minimal hardware overhead.

\subsubsection{Security Analysis of Round Key Based OTP Generation} \label{sec:arch:bandwidth_aware:security_analysis}

The security analysis operates under the standard assumption that AES functions as a pseudorandom permutation (PRP)~\cite{bellare2000authenticated}, and consequently, the AES-CTR mode provides pseudorandom function (PRF) security~\cite{mcgrew2004security}. Under this cryptographic model, the keystream generated by AES-CTR is computationally indistinguishable from a truly random sequence to any polynomial-time adversary. The AES key expansion algorithm is designed as a cryptographically secure key derivation mechanism. According to the AES specification~\cite{daemen2002design}, each round key is derived through a combination of byte substitution, rotation, and XOR operations with round constants, ensuring that round keys exhibit strong avalanche properties and statistical independence. This follows from the fundamental property that XORing a pseudorandom value with any fixed unknown value yields a pseudorandom output. Previous cryptanalytic studies~\cite{biryukov2009related, dunkelman2010improved} have demonstrated the existence of related-key attacks in specific contrived scenarios. These scenarios assume an adversary can manipulate key relationships. However, such attacks are inapplicable to our design, for three key reasons. First, the base key remains secret throughout all operational processes. Second, round keys are used for OTP generation rather than direct encryption. Third, each generated OTP integrates unique per-block metadata (PA, VN), which effectively breaks potential cryptographic correlations.

\subsection{Multi-Level Authentication Mechanism} \label{sec:arch:multi_level_auth}

The granularity of integrity verification presents a critical design trade-off. Overly fine-grained protection inflates security metadata and off-chip memory traffic, degrading performance. Conversely, while coarse-grained approaches reduce security metadata overhead, they can introduce redundant integrity computations, particularly with overlapping data tiles. To resolve this dilemma, we introduce a multi-level integrity verification mechanism that synergizes the benefits of both scales. It leverages the precision of fine-grained verification to eliminate redundant security operations while capturing the low-overhead advantages of coarse-grained protection, thereby substantially reducing or even eliminating off-chip memory accesses for integrity checks. \label{rebuttal:near_zero_overhead}

\subsubsection{Analysis of High-Cost Off-Chip Memory Access in Integrity Checks} \label{sec:arch:multi_level_auth:analysis}

While recent research efforts~\cite{hua2022mgx, lee2022tnpu} have successfully eliminated the off-chip memory access overhead of VNs and Merkle Trees, the overhead introduced by MACs still remains to be adequately addressed. This overhead is exacerbated by both intra-layer and inter-layer data dependencies, which manifest as overlapping computational tiles. Such overlaps lead to redundant data reads and repetitive authentication operations, thereby degrading overall performance. Securator~\cite{shrivastava2023securator} proposed a layer-level freshness and integrity check method to reduce the MACs overhead by XORing MACs of all blocks within a layer, with a block size granularity of 32 bytes. However, this approach has two key limitations. First, it fails to account for tile overlaps within a single layer, leading to redundant integrity computations and unnecessary processing costs. Second, it overlooks the fact that tiling strategies often differ between consecutive layers. As illustrated in Fig.~\ref{figures:arch_framework}(b), the $ofmap$ of a producer layer and the $ifmap$ of its consumer layer may employ distinct tiling patterns that vary in size and dimension. Neglecting these inter-layer dependencies can not only incur additional computational overhead, but also jeopardizes the integrity verification process, potentially leading to erroneous model execution.

\subsubsection{Security Threat Assessment of XOR-based MAC Authentication Scheme} \label{sec:arch:multi_level_auth:threat}

\begin{figure}[t]
    \centering
    \includegraphics[width=0.95\linewidth]{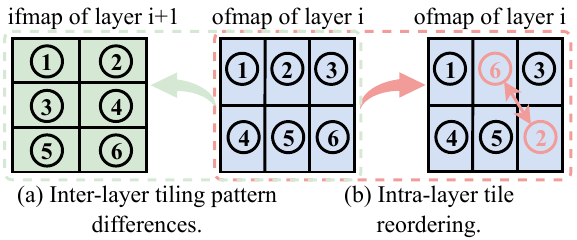}
    \caption{Vulnerability analysis in conventional XOR-based authentication schemes.}
    \label{figures:arch_repa_example}
    \vspace{-10pt}
\end{figure}

\begin{algorithm}[t]
	\DontPrintSemicolon
	\SetKwInOut{Input}{\textbf{Input}}	
	\SetKwInOut{Output}{\textbf{Output}}
	\SetKwFunction{Attack}{\textbf{Attack of RePA}}{}{}
	\SetKwFunction{Defense}{\textbf{Defense of RePA}}{}{}
    \let\oldnl\nl
	\newcommand{\nonl}{\renewcommand{\nl}{\let\nl\oldnl}}

	\Input{ $mac\_set$: MACs corresponding to all tiles within a given layer.}
	\Output{$plaintext\_e$: error plaintext.}
        $layer\_mac \leftarrowtail \sum \oplus mac\_set$ \label{alg:repa:attack:sum_mac}\\
        $mac\_set\_sf \leftarrowtail \textsc{ShuffleOrder}(mac\_set)$ \label{alg:repa:attack:shuffle} \\
        $layer\_mac\_sf \leftarrowtail \sum \oplus mac\_set\_sf$ \\ \label{alg:repa:attack:shuffle_sum}
        $ result \leftarrowtail \textsc{VerifyInteg}(layer\_mac, layer\_mac\_sf)$ \\ \label{alg:repa:attack:result}
        \If {TRUE == result} { \label{alg:repa:attack:verify}
            \For{each encrypted data block ($block\_cipher$) of one layer} { \label{alg:repa:attack:for:begin}
                $plain\_block \leftarrowtail \textsc{Decrypt}(cipher\_block)$ \label{alg:repa:attack:for:end}
            }
	}
	\caption{Security threat of \textsc{RePA}} \label{alg:repa:attack}
\end{algorithm}

The XOR-based MAC scheme offers parallelizability, incrementality, and provable security on par with chaining MAC~\cite{bellare1995xor}, enabling layer-level authentication with a single aggregated MAC instead of per-tile MACs. This significantly reduces off-chip memory overhead. However, applying this scheme to DNN layers presents two critical challenges. First, varying tiling patterns across layers cause inter-layer dependency mismatches. Naive XOR-based aggregation across these boundaries produces false positives, where legitimate data is erroneously flagged as tampered (Fig.~\ref{figures:arch_repa_example}(a)). Second, the XOR-aggregation scheme enables the \underline{Re}-\underline{P}ermutation \underline{A}ttack (\textsc{RePA}), detailed in Algorithm~\ref{alg:repa:attack} (lines~\ref{alg:repa:attack:sum_mac}-\ref{alg:repa:attack:for:end}). This attack exploits XOR's commutative property to reorder tiles within a layer without detection. Since XOR aggregation is order-independent, the permuted layer MAC ($layer\_mac\_sf$) equals the original MAC ($layer\_mac$), bypassing integrity verification (line~\ref{alg:repa:attack:result}). Although individual tiles decrypt correctly, their altered spatial sequence violates DNN structural dependencies, producing erroneous inference results. As Fig.~\ref{figures:arch_repa_example}(b) shows, swapping tile-2 and tile-6 corrupts the layer structure but evades naive XOR-based checks, resulting in a false negative that compromises model output.

In contrast to SecureLoop's analytical paradigm~\cite{lee2023secureloop}, which utilizes linear congruence modeling and simulated annealing, our proposed the greatest common divisor (GCD)-based computational framework to define optimal authentication blocks ($opt\_blk$). As detailed in Algorithm~\ref{alg:gcd}, it intrinsically eliminates redundant off-chip memory access for security metadata by partitioning tiles into $opt\_blk$. The procedure iteratively processes each dimension of the data structure (line~\ref{alg:gcd:for}). Leveraging the deterministic nature of DNN computation, the size of the overlapped region between two tiling patterns can be precisely calculated (line~\ref{alg:gcd:calcoverlapsize}). These overlaps are typically caused by inter-layer dependencies or intra-layer reordering. Subsequently, the GCD is computed for the sizes of this overlapped and the remaining non-overlapped region to determine the optimal block size for that specific dimension (line~\ref{alg:gcd:calcgcd}). By iteratively applying this process across all dimensions, the final $opt\_blk$ is constructed from the set of all calculated dimensional block sizes (line~\ref{alg:gcd:optblk}). Table~\ref{table:macs} provides a detailed comparison of the features among these four types of MACs: $opt\_blk\_mac$, $tile\_mac$, $layer\_mac$, and $model\_mac$.

\begin{algorithm}[t]
	\DontPrintSemicolon
	\SetKwInOut{Input}{\textbf{Input}}	
	\SetKwInOut{Output}{\textbf{Output}}
    \let\oldnl\nl
	\newcommand{\nonl}{\renewcommand{\nl}{\let\nl\oldnl}}

	\Input{ $n$: Number of dimensions; $l^i_1, l^i_2$: length of two different tiling patterns in dimension $i$.}
	\Output{$opt\_blk$: optimal authentication block.}

    \For{ $i \leq n$ } { \label{alg:gcd:for}
        $l^i_{12} \leftarrowtail \textsc{CalcOverlapSize}(l^i_1, l^i_2)$ \\ \label{alg:gcd:calcoverlapsize}
        $opt\_blk^i \leftarrowtail \textsc{CalcGCD}(l^i_1 - l^i_{12}, l^i_2 - l^i_{12}, l^i_{12})$ \\ \label{alg:gcd:calcgcd}
    }
    $opt\_blk \leftarrowtail (opt\_blk^1, opt\_blk^2, \dots, opt\_blk^n)$ \\ \label{alg:gcd:optblk}
	\caption{GCD-based optimal authentication block computational framework} \label{alg:gcd} 
    \label{alg:gcd_framework}
\end{algorithm}

\begin{table}[t]
\caption{Comparison of characteristics across multi-level authentication mechanisms.}
\label{table:macs}
\centering
    \footnotesize
    \renewcommand{\arraystretch}{0.85}
    \setlength{\tabcolsep}{1.8pt}
    \begin{tabular}{c|c|c|c}
        \toprule
        \multicolumn{1}{c|}{\textbf{Granularity}}   & \multicolumn{1}{c|}{\textbf{Flexibility}} & \multicolumn{1}{c|}{\textbf{Off-chip Access Overhead}} & \multicolumn{1}{c}{\textbf{Storage}} \\ 
        \midrule                        
        $opt\_blk\_mac$   & \begin{tikzpicture} \fill[gray!100] (0,0) circle (0.15cm); \end{tikzpicture}   &  \begin{tikzpicture} \draw (0,0) circle (0.15cm); \fill[orange!50] (0,0) -- (0.15,0) arc (0:60:0.15cm) -- cycle; \end{tikzpicture} & Off-chip\\
        \rowcolor[HTML]{EFEFEF}
        $tile\_mac$        & \begin{tikzpicture} \fill[gray!90] (0,0) circle (0.15cm); \end{tikzpicture}   & \begin{tikzpicture}  \draw (0,0) circle (0.15cm); \fill[orange!50] (0,0) -- (0.15,0) arc (0:40:0.15cm) -- cycle; \end{tikzpicture} & Off/On-chip \\
        $layer\_mac$        & \begin{tikzpicture} \fill[gray!60] (0,0) circle (0.15cm); \end{tikzpicture}   & \begin{tikzpicture}  \draw (0,0) circle (0.15cm); \fill[orange!50] (0,0) -- (0.15,0) arc (0:20:0.15cm) -- cycle; \end{tikzpicture} & Off/On-chip \\
        \rowcolor[HTML]{EFEFEF}
        $model\_mac$       &  \begin{tikzpicture} \fill[gray!30] (0,0) circle (0.15cm); \end{tikzpicture} &   \begin{tikzpicture} \draw (0,0) circle (0.15cm); \fill[orange!50] (0,0) -- (0.15,0) arc (0:0:0.15cm) -- cycle; \end{tikzpicture}   &  On-chip \\
        \bottomrule
    \end{tabular}
    \begin{flushleft}
        \small
         *Color depth indicates intensity, while coverage area indicates ratio.
    \end{flushleft}
    \vspace{-10pt}
\end{table}

\subsubsection{Proposed Solution} \label{sec:arch:multi_level:solution}

\begin{figure}
    \centering
    \includegraphics[width=0.98\linewidth]{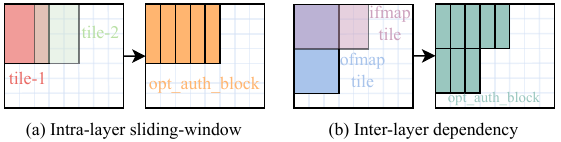}
    \caption{Illustrative examples of optimal authentication block assignment strategies.}
    \label{figures:arch_repa_overlaps}
    \vspace{-10pt}
\end{figure}

Fig.~\ref{figures:arch_repa_overlaps} provides examples illustrating the derivation of optimal authentication block assignment strategies using the proposed GCD-based computational framework. In Fig.~\ref{figures:arch_repa_overlaps}(a), the intra-layer scenario demonstrates how $opt\_blk$ is determined while accounting for overlaps induced by the intra-layer sliding window mechanism. In this case, since all tiles within the current intra-layer share a uniform tiling pattern, alignment optimization is confined to the horizontal dimension. In Fig.~\ref{figures:arch_repa_overlaps}(b), the inter-layer scenario highlights the consideration of overlaps arising from inter-layer dependencies, which necessitate alignment optimization across both horizontal and vertical dimensions due to the emergence of heterogeneous tiling patterns. Notably, this methodology is inherently extensible to higher-dimensional spaces for solving $opt\_blk$, where alignment optimization can be generalized to incorporate additional spatial dimensions as required.

\begin{algorithm}[t]
    \DontPrintSemicolon
    \SetKwInOut{Input}{\textbf{Input}}	
    \SetKwInOut{Output}{\textbf{Output}}
    \SetKwFunction{Attack}{\textbf{Attack of RePA}}{}{}
    \SetKwFunction{Defense}{\textbf{Defense of RePA}}{}{}
    \let\oldnl\nl
    \newcommand{\nonl}{\renewcommand{\nl}{\let\nl\oldnl}}

    \Input{$opt\_blk, layer\_id, opt\_blk\_idx$: optimal authentication encrypted data block, layer number, and block index of $layer\_id$. \\ }
    \Output{$layer\_mac$: a layer MAC.\\}

    \For{each $opt\_blk$ of $layer\_id$} { \label{alg:arch_repa_defense_for_begin}
        $opt\_blk\_mac \leftarrowtail \textsc{Auth}_{K_h}(opt\_blk$ $||$ PA $||$ VN $||$ $layer\_id$ $||$ $opt\_blk\_idx$)  \label{alg:arch_repa_defense_for_auth}  \\     
        $layer\_mac \leftarrowtail layer\_mac \oplus opt\_blk\_mac$ \label{alg:arch_repa_defense_for_end}
    }
    \caption{Defense against \textsc{RePA}} \label{alg:arch_repa_defense}
\end{algorithm}

To defend against \textsc{RePA} attacks targeting XOR-based schemes, lines~\ref{alg:arch_repa_defense_for_begin}-\ref{alg:arch_repa_defense_for_end} of Algorithm~\ref{alg:arch_repa_defense} introduce corresponding countermeasures. For each $opt\_blk$ in layer $layer\_id$, we compute $opt\_blk\_mac$ via the \textsc{Auth} function incorporating PA, VN, $layer\_id$, and a unique index $opt\_blk\_idx$. This cryptographic binding breaks XOR's commutativity, the core vulnerability of \textsc{RePA}, by making MAC computation depend on both block content and its fixed position index. Even if an adversary permutes $opt\_blk$ ordering, the invariant metadata (PA, VN, $layer\_id$, $opt\_blk\_idx$) enables tamper detection, causing integrity verification to fail. This hierarchical design allows $layer\_mac$ storage in on-chip SRAM, eliminating off-chip memory access overhead. Similarly, $model\_mac$ represents entire model weights on-chip with a single MAC, further reducing off-chip costs while conserving SRAM. Verification results become available only at inference completion. In summary, our multi-level integrity verification mechanism eliminates off-chip metadata overhead while maintaining comprehensive security guarantees.

\subsubsection{Security Analysis of Multi-Level Authentication Mechanism} \label{sec:arch:multi_level_auth:security_analysis}

Within the threat model defined in Section~\ref{section:threat_moti}, where adversaries can monitor off-chip memory traffic but cannot observe on-chip operations, the multi-level authentication mechanism presents controlled information leakage characteristics. This hierarchical approach enables authentication metadata to be stored in secure on-chip SRAM. Even when aggregated $layer\_mac$ values require off-chip storage, they generate only minimal memory traffic. Since adversaries cannot access on-chip data and can only observe limited off-chip MAC traffic, this fundamentally restricts the attack surface available for traffic analysis, thereby enhancing security. Furthermore, as analyzed in Section~\ref{sec:arch:bandwidth_aware:security_analysis}, the XOR operations preserve pseudorandomness without compromising cryptographic security guarantees. While adversaries may attempt to infer structural information via timing or power side-channels, DNN model architecture inference falls outside the scope of our security objectives, as noted in Section~\ref{section:threat_moti}. Even if model structures are deduced, this knowledge offers no advantage in compromising data confidentiality or integrity without the secret key.

\section{Experimental Evaluation} \label{section:evaluation}

\subsection{Experimental Setup}

\subsubsection{Simulation Methodology}
To comprehensively assess inference behaviours of DNN accelerators, we employ SCALE-Sim2~\cite{samajdar2018scale,samajdar2020systematic}, an open-source cycle-level DNN simulator developed by ARM Research. This simulator serves a dual-purpose in our experimental methodology: first, it enables in-depth investigations into the inference execution dynamics of diverse DNN models, ranging from lightweight architectures suitable for edge devices to computationally intensive networks designed for high-performance computing environments. Second, it facilitates the meticulous analysis of performance overheads associated with different memory protection schemes, which is crucial for understanding the trade-offs between security and computational efficiency.

The custom-designed DNN accelerator, integrated with the simulation framework, provides fine-grained insights into the operational characteristics of the systolic array. Concurrently, it captures comprehensive DRAM access traces. Upon obtaining the raw DRAM access traces, we apply a suite of memory protection mechanisms, each tailored to specific security requirements. By integrating these protection schemes into the simulation workflow, we accurately compute the execution time and bandwidth usage metrics. This process generates augmented DRAM traces that reflect the impact of security measures on memory access patterns. Finally, to validate the performance implications of these modified DRAM access traces, we utilize Ramulator2~\cite{luo2023ramulator}, a highly accurate DRAM simulator. Ramulator2 models the complex behavior of modern DRAM architectures, enabling us to simulate the total DRAM access patterns with high fidelity. Additionally, we employ DRAMsim3~\cite{li2020dramsim3} to perform detailed energy consumption analysis on memory access operations under various memory protection schemes. The memory configuration adopts DDR4\_16Gb\_x8\_3200, which specifies a 16Gb DDR4 DRAM device and a data rate of 3200 MB/s.

\subsubsection{Accelerator Configurations} 
Table~\ref{table:eval_configs} lists the configurations of DNN accelerators, including a server NPU (Google TPU v1) and an edge NPU (Samsung Exynos 990), both of which employ the output stationary dataflow. To achieve a balance between DNN computation and memory bandwidth, we configured four 64-bit DDR channels in our simulations for both the server and edge NPUs.

\begin{table}[t]
    \caption{DNN accelerator configurations~\cite{jouppi2017datacenter,song20197}.}
    \label{table:eval_configs}
    \centering
    \footnotesize
    \renewcommand{\arraystretch}{0.85}
    \setlength{\tabcolsep}{3pt}
    \begin{tabular}{c|c|c}
        \toprule
        \textbf{Metrics} & \multicolumn{1}{c|}{\textbf{Server (Google TPU v1)}}   & \multicolumn{1}{c}{\textbf{Edge (Samsung Exynos 990)}} \\ 
        \midrule                           
        \textbf{PE}              & 256 $\times$ 256  in systolic array   & 32 $\times$ 32 in systolic array \\
        \rowcolor[HTML]{EFEFEF}
        \textbf{Bandwidth}        & 20 GB/s with 4 channels   & 10 GB/s with 4 channels \\
        \textbf{Frequency}        &  1 GHz &   2.75 GHz   \\
        \rowcolor[HTML]{EFEFEF}
        \textbf{SRAM}             & 24 MB & 480 KB \\
        \textbf{Precision}        & 1-B for per element  & 1-B for per element \\
        \textbf{Dataflow}         & output stationary & output stationary \\
        \bottomrule
    \end{tabular}
    \vspace{-10pt}
\end{table}

\subsubsection{Memory Protection Implementation}
We have developed secure DNN accelerators that incorporate Intel SGX, MGX, and our proposed security mechanisms. An unprotected baseline accelerator serves as a reference benchmark, allowing for direct comparison of performance and security trade-offs. In our experimental setup, the protected memory size is standardized at 16GB to ensure consistency across different configurations. For the Intel SGX implementation, we utilize a multi-level integrity tree architecture. This structure features 56-bit VNs and 64-bit MACs and includes a 16KB VN cache and an 8KB MAC cache, both employing a Least Recently Used (LRU) replacement policy for write-back and write-allocate strategies. Our evaluation considers two protection granularities: 64B and 512B. This variation enables us to analyze the impact of different granularity levels on security robustness and performance overhead. To maintain fairness in the experimental comparison, layer MACs are consistently stored in off-chip memory.

\subsubsection{Cross-domain Workloads} 
To thoroughly evaluate the proposed secure DNN accelerator framework, we conduct comprehensive assessments across a diverse array of DNN models. Our evaluation suite encompasses Lenet (let), Alexnet (alex), Mobilenet (mob), ResNet18 (rest), GoogleNet (goo), DLRM (dlrm), AlphaGoZero (algo), DeepSpeech2 (ds2), FasterRCNN (fast), NCF\_recommendation (ncf), Sentimental\_seqCNN (sent), Transformer\_fwd (trf), Yolo\_tiny (yolo). These models are strategically selected from various machine learning domains, spanning computer vision, speech recognition, natural language processing, gaming, and personalized recommendation, to ensure their effectiveness in real-world scenarios.

\begin{table*}[t]
    \caption{Comparison of memory protection schemes.}
    \label{table:comparison}
    \centering
    \footnotesize
    \renewcommand{\arraystretch}{0.85}
    \setlength{\tabcolsep}{4pt}
    \begin{tabular}{c|c|c|c|c|c}
        \toprule
        \textbf{Protection Scheme} & \textbf{Encryption Granularity}  & \textbf{Integrity Granularity} 
        & \textbf{Off-chip Memory Access}     & \textbf{DNN Tiling Pattern}   &  \textbf{Encryption Scalability}  \\ 
        \midrule
                        
        \textbf{SGX-64B} & 16B & 64B & MAC,VN,IT  & \textcolor{pink}{\large \ding{55}}  &  \textcolor{pink}{\large \ding{55}} \\
        \rowcolor[HTML]{EFEFEF}
        \textbf{SGX-512B} & 16B & 512B & MAC,VN,IT & \textcolor{pink}{\large \ding{55}} &  \textcolor{pink}{\large \ding{55}} \\
        \textbf{MGX-64B} & 16B & 64B & MAC &\textcolor{pink}{\large \ding{55}} & \textcolor{pink}{\large \ding{55}} \\
        \rowcolor[HTML]{EFEFEF}
        \textbf{MGX-512B} & 16B & 512B & MAC & \textcolor{pink}{\large \ding{55}} & \textcolor{pink}{\large \ding{55}} \\
        \rowcolor[HTML]{DAE8FC}
        \textbf{Ours} & bandwidth-aware & multi-level & minimal to no cost & \textcolor{green}{\large \cmark} & \textcolor{green}{\large \cmark} \\

        \bottomrule
    \end{tabular}

    \begin{flushleft}
        \small
        \hspace{0.6 cm} * ``IT'' signifies integrity tree.
    \end{flushleft}
    \vspace{-10pt}
\end{table*}

\subsection{Experimental Results}
We conduct a comprehensive comparative analysis of memory traffic overhead and computational performance across an unsecure baseline and five distinct protection schemes: SGX-64B, SGX-512B, MGX-64B, MGX-512B, and our proposed approach, as detailed in Table~\ref{table:comparison}. Memory traffic represents the volume of data exchanged between off-chip memory and accelerators, serving as a critical metric for evaluating system efficiency. All experimental results are normalized to the unsecure baseline to facilitate direct comparison across different protection mechanisms.

\subsubsection{Scalabilities of Area and Power}

\begin{figure}[t]
    \centering
    \includegraphics[width=0.45\linewidth]{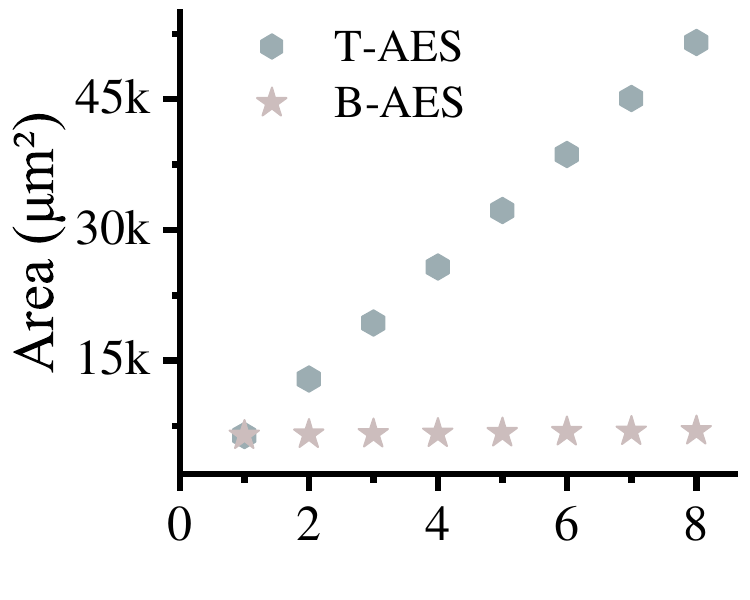} \hfil
    \includegraphics[width=0.45\linewidth]{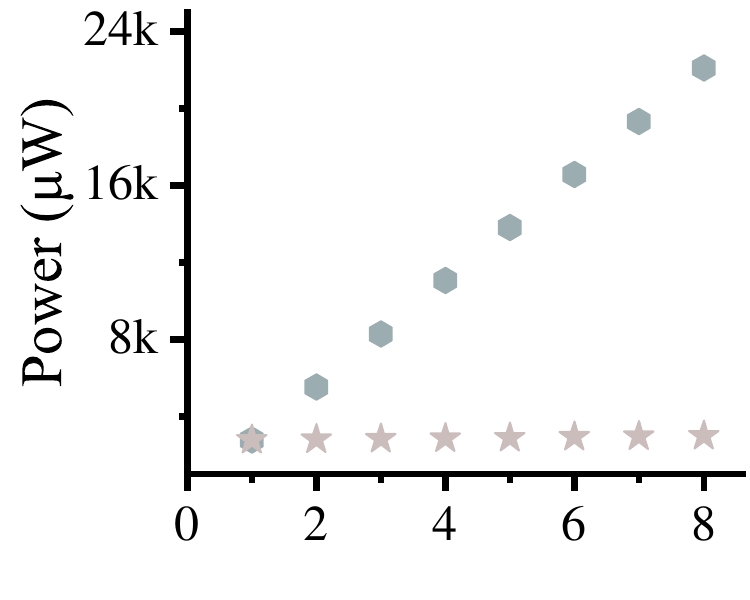}
    \caption{The area and power with increasing AES engine bandwidth requirements.}
    \label{figures:eval_area_power}
    \vspace{-10pt}
\end{figure}

To systematically evaluate the hardware cost implications of different cryptographic approaches, we developed a comprehensive analysis based on 28nm technology, leveraging the well-established AES engine implementations reported in~\cite{banerjee2017energy}. In our comparative analysis, we distinguish between two fundamentally different architectural approaches: the bandwidth-aware cryptographic scheme (denoted as B-AES) and prior approaches that rely on multiple parallel AES engines (referred to as T-AES). Through extensive simulation studies, we examine how hardware area and power consumption scale with increasing cryptographic engine bandwidth requirements. As clearly demonstrated in Fig.~\ref{figures:eval_area_power}, the x-axis represents the normalized cryptographic engine bandwidth requirements (expressed as multiples of 16B, the throughput of a single AES engine), while the y-axis depicts the corresponding hardware area (left panel) and power consumption (right panel). The experimental results reveal a striking contrast between the two approaches. Specifically, the T-AES method exhibits substantial linear growth in both area and power consumption as bandwidth requirements increase, while our proposed B-AES maintains remarkably consistent hardware costs, demonstrating near-constant area and power consumption regardless of bandwidth scaling requirements. This scalability characteristic highlights the fundamental efficiency advantage of our bandwidth-aware approach, making it particularly well-suited for high-performance DNN accelerators where resource constraints are critical design considerations.

\subsubsection{Effect of Configurations} \label{cover:eval_configuration}

\begin{figure}
    \centering
    \includegraphics[width=0.7\linewidth]{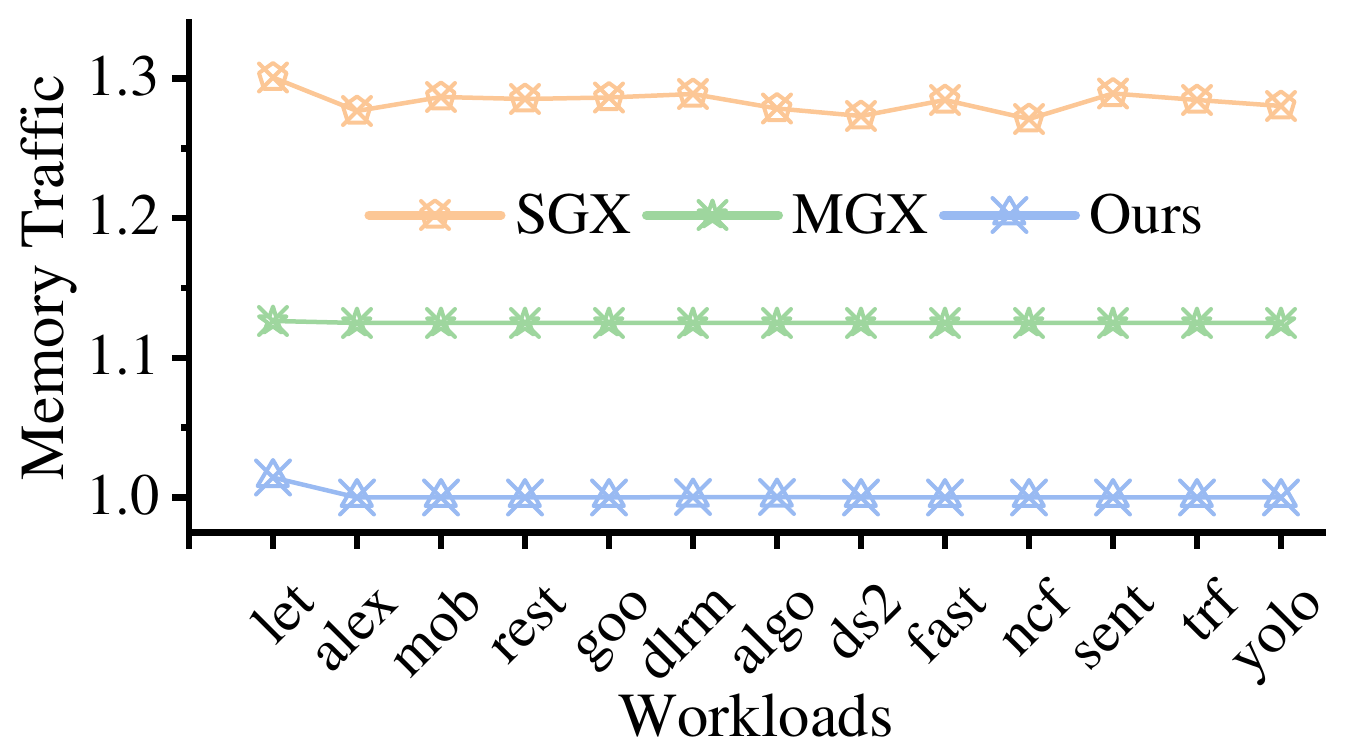}
    \caption{Memory traffic for secure accelerators varies across configurations.}
    \label{figures:eval_edge_server_traffic}
    \vspace{-10pt}
\end{figure}

As clearly illustrated in Fig.~\ref{figures:eval_edge_server_traffic}, the experimental results reveal distinct performance characteristics across the three protection approaches. First, SGX consistently demonstrates the highest memory traffic overhead across all evaluated workloads, maintaining approximately $1.30 \times$ normalized memory traffic. This substantial overhead stems primarily from SGX's simultaneous adoption of an integrity tree structure for replay attack defense and block-based Message Authentication Codes (MACs), consequently causing SGX's memory traffic to surge by nearly $30\%$ compared to the unprotected baseline across diverse workload patterns. In contrast, MGX demonstrates significantly improved efficiency by maintaining approximately $1.13 \times$ normalized memory traffic across all workloads. By strategically leveraging the deterministic access characteristics inherent in DNN computations, MGX enables real-time on-chip computation of VNs, thereby effectively eliminating the substantial overhead introduced by the integrity tree structure and yielding approximately $13\%$ reduction in memory traffic compared to SGX. Most remarkably, our proposed framework achieves near-optimal performance by maintaining approximately equivalent memory traffic, essentially matching the unprotected baseline performance. This exceptional efficiency is attributed to our innovative multi-level authentication mechanism that fundamentally addresses the block-based MAC overhead challenge through exclusively employing layer-based MACs for integrity verification, which, due to their minimal storage footprint, can be stored entirely on-chip, effectively eliminating costly off-chip memory accesses for security metadata retrieval and resulting in negligible memory traffic overhead regardless of workload complexity or characteristics.

\subsubsection{Performance}

\begin{figure}[t]
	\centering
	\includegraphics[width=0.8\linewidth]{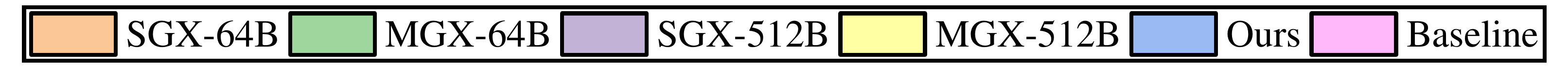}\\
	\subfloat[Server NPU]{\includegraphics[width= 0.98\linewidth]{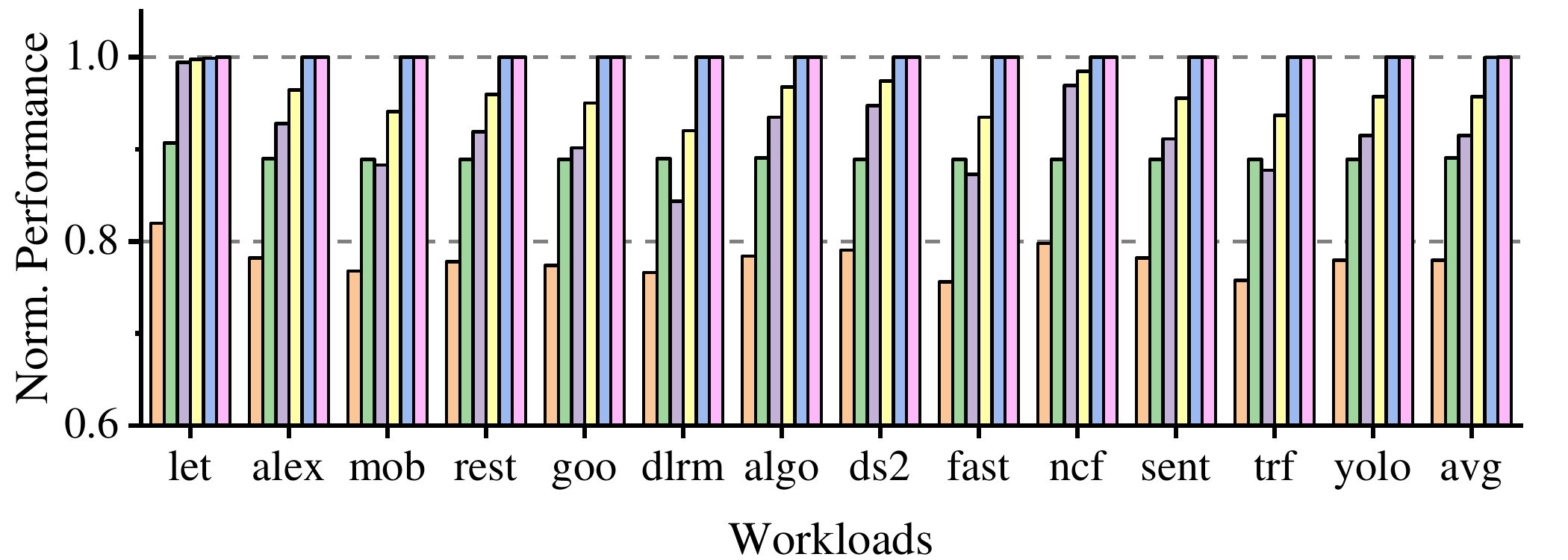}}\\
	\subfloat[Edge NPU]{\includegraphics[width= 0.98\linewidth]{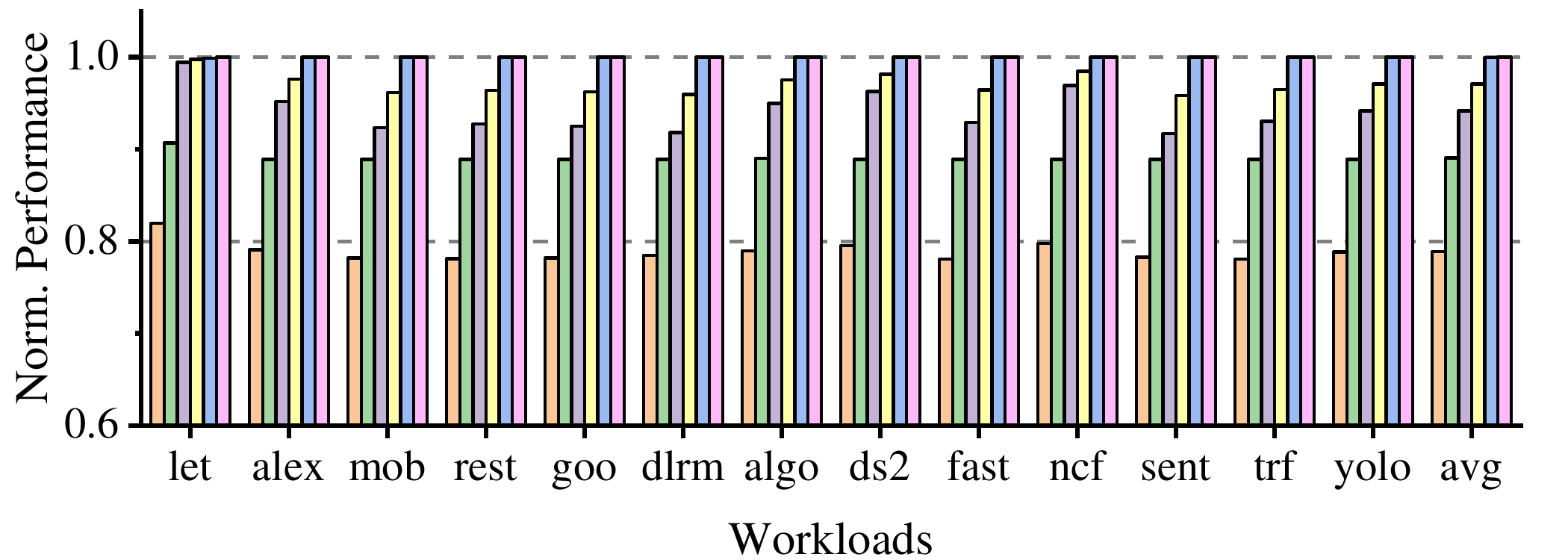}}
	\caption{The normalized performance of memory protection schemes for various workloads.}
    \label{figures:eval_performance_os}
    \vspace{-10pt}
\end{figure}

Fig.~\ref{figures:eval_performance_os} reveals the computational overhead characteristics of various memory protection schemes across both device architectures. A consistent performance hierarchy emerges where prior protection mechanisms impose significant computational burdens while our proposed solution maintains exceptional efficiency. For Server NPU, Fig.~\ref{figures:eval_performance_os}(a) demonstrates that SGX-64B incurs the most substantial performance penalty with $22.04\%$ execution slowdown, followed by MGX-64B with $10.93\%$ overhead, SGX-512B with $8.49\%$ degradation, and MGX-512B achieving $4.28\%$ performance reduction compared to the unprotected baseline. Conversely, our proposed framework delivers remarkable computational efficiency with negligible performance impact, effectively eliminating the security-performance trade-off. This superior performance originates from our multi-level integrity verification architecture that resolves the computational bottleneck in off-chip security metadata management—through strategic deployment of compact layer-based MACs stored directly within on-chip SRAM, we eliminate the performance penalties associated with frequent off-chip memory accesses for integrity verification. The Edge NPU evaluation, presented in Fig.~\ref{figures:eval_performance_os}(b), substantiates these trends with consistent results: SGX-64B, MGX-64B, SGX-512B, and MGX-512B exhibit performance degradations of $21.10\%$, $10.95\%$, $5.84\%$, and $2.90\%$ respectively, while our proposed solution maintains virtually undetectable performance overhead, demonstrating the universal effectiveness of our architectural innovations across heterogeneous computational platforms.

\subsubsection{Memory Traffic}

\begin{figure}[t]
	\centering
    \includegraphics[width=0.8\linewidth]{figures/eval_legend.pdf}\\
	\subfloat[Server NPU]{
        \includegraphics[width= 0.98\linewidth]{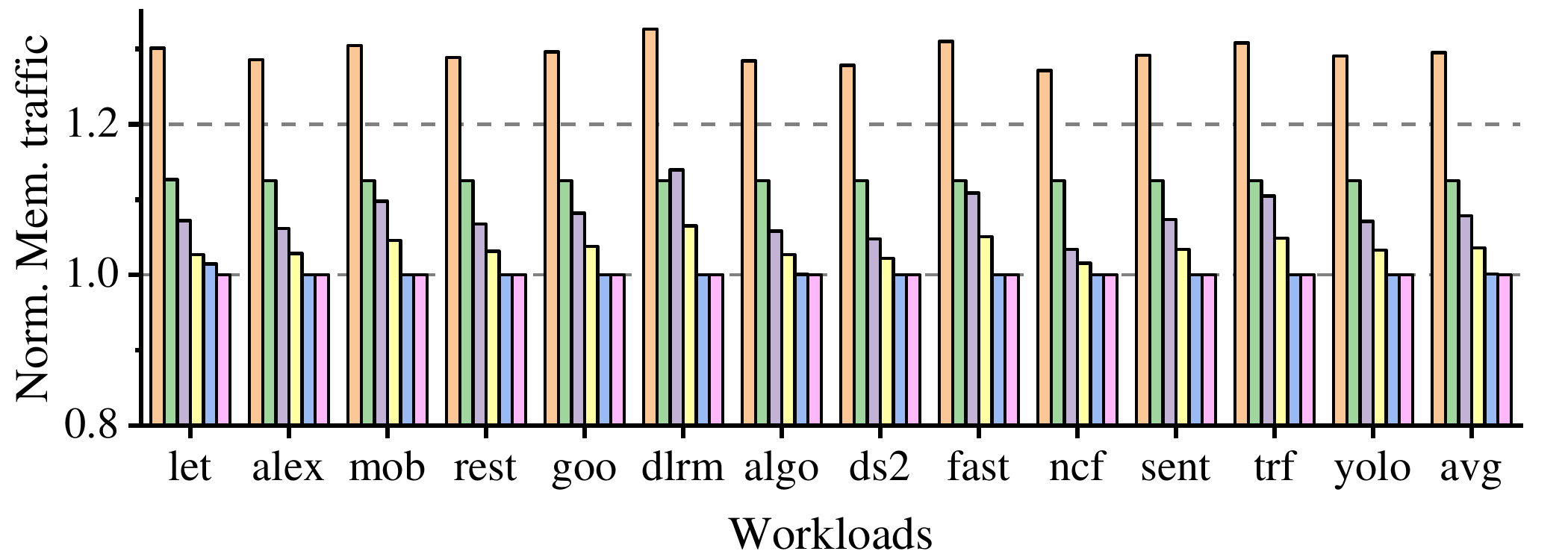} 
        }\\
	\subfloat[Edge NPU]{
		\includegraphics[width= 0.98\linewidth]{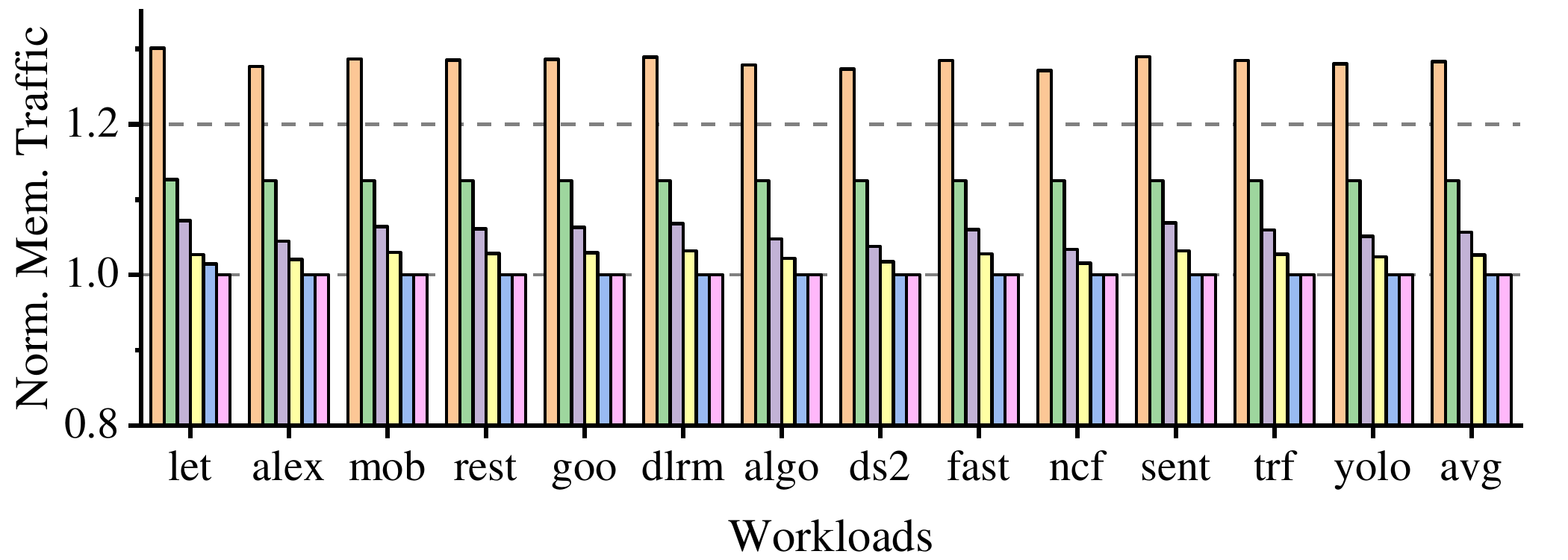}
        }
	\caption{The normalized memory traffic of memory protection schemes for various workloads.}
    \label{figures:eval_traffic_os}
    \vspace{-10pt}
\end{figure}

Building upon the configuration analysis presented earlier, Fig.~\ref{figures:eval_traffic_os} provides a comprehensive quantitative evaluation of memory traffic overhead using the DNN simulation configurations detailed in Table~\ref{table:eval_configs}. The experimental results demonstrate significant performance variations across different protection schemes and device configurations. For the baseline SGX implementations, the data reveals substantial memory traffic increases: SGX-64B elevates memory traffic by $30\%$ for Server NPU and $28.29\%$ for Edge NPU on average. In contrast, MGX-64B demonstrates considerably improved efficiency by eliminating the overhead from additional VNs and MTs, resulting in traffic increases of only $12.51\%$ and $12.63\%$, respectively. This substantial improvement underscores the effectiveness of leveraging DNN-specific access patterns for security optimization. Furthermore, the analysis reveals that increasing the protection granularity from 64B to 512B yields significant memory traffic reductions across all protection schemes. Specifically, SGX-512B achieves traffic reductions of $7.83\%$ on Server NPU and $5.13\%$ on Edge NPU compared to SGX-64B. Similarly, MGX-512B demonstrates reductions of $3.59\%$ and $2.39\%$ on these respective devices. While expanding the protection granularity to 512 bytes effectively reduces memory traffic by minimizing security metadata overhead, this approach presents trade-offs. The adoption of larger data blocks can potentially create inefficiencies due to misalignment with intra-layer tiling overlaps and varying inter-layer tiling patterns, which may hinder optimal memory access patterns and resource utilization, emphasizing the critical need for balanced granularity selection. Most remarkably, our proposed approach employs an innovative multi-level integrity verification mechanism that aggregates all block-based MACs through XOR operations into a unified layer MAC. As demonstrated in Fig.~\ref{figures:eval_traffic_os}, our solution achieves exceptional efficiency with only $0.12\%$ overhead for Server NPU and $0.03\%$ for Edge NPU. This outstanding performance clearly highlights the superiority of our proposed protection mechanism over existing state-of-the-art approaches, effectively bridging the gap between security and performance requirements.

\subsubsection{Energy Efficiency} \label{cover:eval_energy}

\begin{figure}[t]
	\centering
	\includegraphics[width=0.8\linewidth]{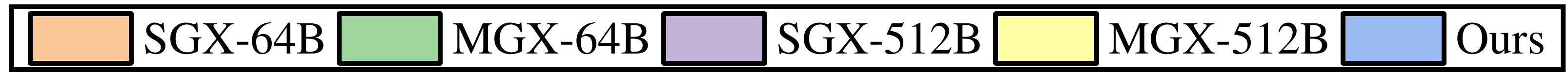}\\
	\subfloat[Server NPU]{\includegraphics[width= 0.98\linewidth]{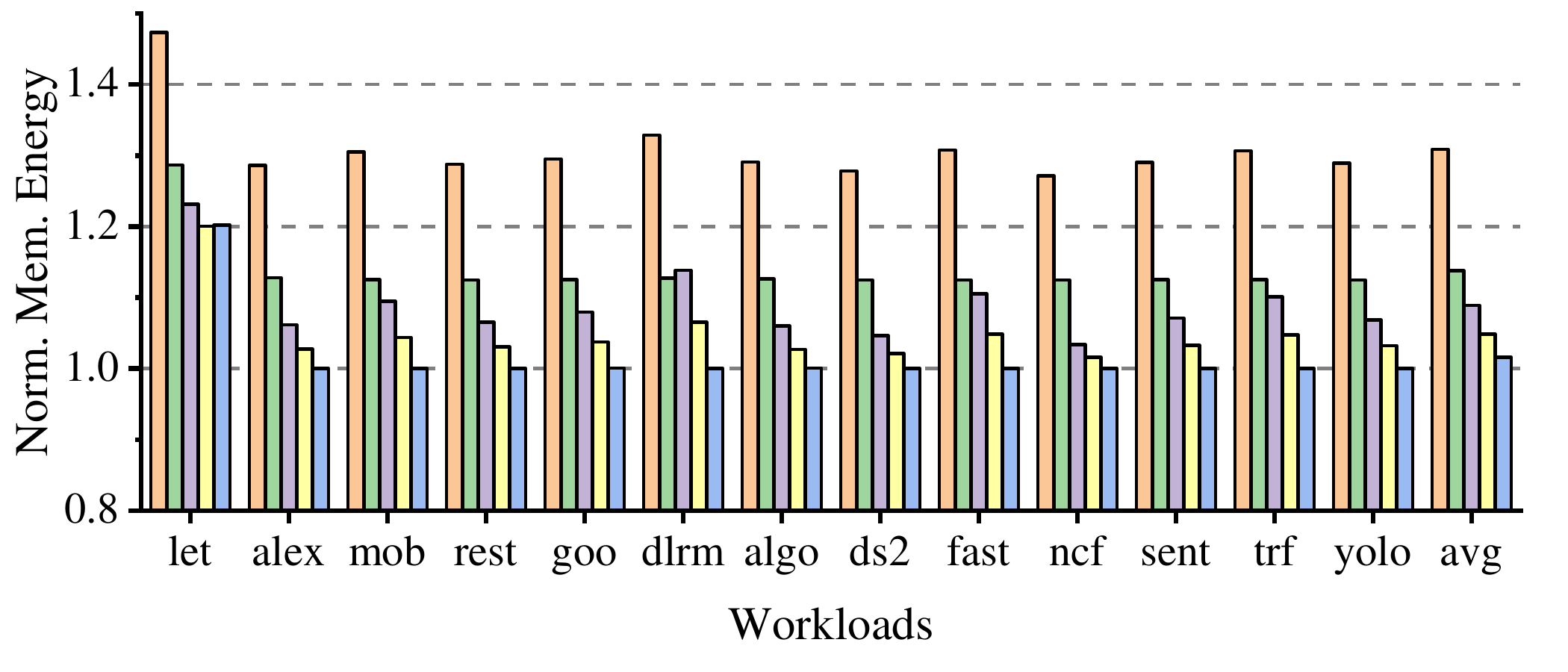}}\\
	\subfloat[Edge NPU]{\includegraphics[width= 0.98\linewidth]{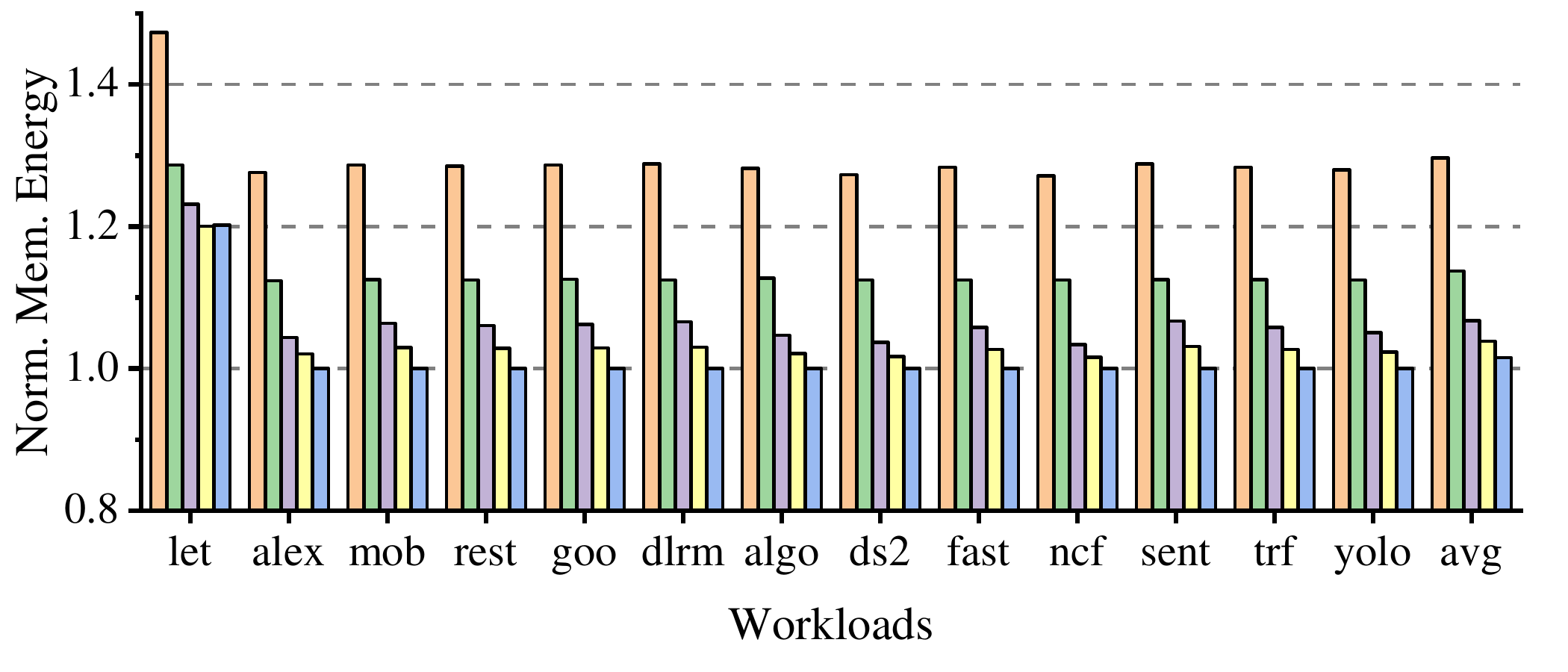}}
	\caption{The normalized memory access energy consumption of memory protection schemes for various workloads.}
    \label{figures:eval_energy_os}
    \vspace{-10pt}
\end{figure}

The energy consumption analysis, as presented in Fig.~\ref{figures:eval_energy_os}, demonstrates that memory protection schemes exhibit energy overhead patterns that directly mirror their respective memory traffic characteristics. Among the evaluated protection mechanisms, SGX implementations impose the most significant energy penalties, with SGX-64B consuming approximately $30.8\%$ and $29.7\%$ additional energy for Server NPU and Edge NPU respectively, primarily due to the extensive off-chip metadata management required for maintaining integrity trees and version numbers. The MGX-64B approach substantially mitigates these energy costs by leveraging on-chip computation capabilities, reducing energy overhead to $13.8\%$ and $12.9\%$ through the strategic elimination of off-chip VN and MT accesses, thereby demonstrating the effectiveness of exploiting DNN-specific computational patterns for energy optimization. Protection granularity scaling from 64B to 512B further enhances energy efficiency across both SGX and MGX implementations: SGX-512B reduces energy penalties to $21.9\%$ and $22.9\%$, while MGX-512B achieves even greater efficiency with only $8.9\%$ and $9.9\%$ overhead, illustrating how reduced metadata volume directly translates to lower energy consumption. Our proposed framework maintains an additional energy consumption of less than $1.6\%$ across all workloads and device configurations. This energy conservation results from our multi-level integrity verification strategy, which eliminates the prior energy-intensive cycle of off-chip security metadata retrieval and storage.

\section{Conclusion} \label{section:conclusion}

In the current technological landscape, where DNNs are increasingly deployed in untrusted environments such as public clouds and edge computing nodes, safeguarding the confidentiality and integrity of data processed by DNN accelerators has emerged as a critical challenge. In this work, we introduce an innovative and groundbreaking secure and efficient DNN accelerator architecture framework, which serves as a robust solution to address these security concerns while maintaining high computational performance. Using bandwidth-aware cryptographic scheme and multi-level authentication mechanism, this framework provides excellent scalability with minimal overhead to match computational bandwidth of accelerators, and significantly reduces or even eliminates off-chip memory access overhead. Experimental results show that the proposed framework meets bandwidth requirements with negligible hardware overhead, and significantly reduces the performance overhead compared to the state-of-the-art approaches. As DNNs become more embedded in critical infrastructure and privacy-sensitive applications, our framework provides a robust foundation for deploying DNN accelerators that balance security rigor with computational efficiency, advancing the development of trustworthy AI hardware architectures.


\bibliographystyle{IEEEtran}
\bibliography{reference}

\end{document}